\documentclass[aps,prl,twocolumn,showpacs,superscriptaddress,floatfix]{revtex4-1}
\usepackage{graphicx}
\usepackage{amsmath}
\usepackage{amssymb}
\usepackage[labeled,resetlabels]{multibib}
\newcites{S}{References}
\usepackage{hyperref}

\newcounter{mysection}
\makeatletter
     \@addtoreset{figure}{mysection}
\makeatother

\DeclareMathOperator{\sech}{sech}

\begin{document}
\title{Precise quantization of anomalous Hall effect near zero magnetic field}

\author{A.\ J.\ Bestwick}
\affiliation{Department of Physics, Stanford University, Stanford, CA 94305, USA}
\affiliation{Stanford Institute for Materials and Energy Sciences, SLAC National Accelerator Laboratory, 2575 Sand Hill Road, Menlo Park, California 94025, USA}

\author{E.\ J.\ Fox}
\affiliation{Department of Physics, Stanford University, Stanford, CA 94305, USA}
\affiliation{Stanford Institute for Materials and Energy Sciences, SLAC National Accelerator Laboratory, 2575 Sand Hill Road, Menlo Park, California 94025, USA}

\author{Xufeng Kou}
\affiliation{Department of Electrical Engineering, University of California, Los Angeles, CA 90095, USA}

\author{Lei Pan}
\affiliation{Department of Electrical Engineering, University of California, Los Angeles, CA 90095, USA}

\author{Kang L.\ Wang}
\affiliation{Department of Electrical Engineering, University of California, Los Angeles, CA 90095, USA}

\author{D.\ Goldhaber-Gordon}
\email[To whom correspondence should be addressed; Email: ]{goldhaber-gordon@stanford.edu}
\affiliation{Department of Physics, Stanford University, Stanford, CA 94305, USA}
\affiliation{Stanford Institute for Materials and Energy Sciences, SLAC National Accelerator Laboratory, 2575 Sand Hill Road, Menlo Park, California 94025, USA}

\pacs{75.47.--m, 73.43.Fj, 75.45.+j, 75.50.Pp}
\date{16 January 2015}

\begin{abstract}
We report a nearly ideal quantum anomalous Hall effect in a three-dimensional topological insulator thin film with ferromagnetic doping. Near zero applied magnetic field we measure exact quantization in Hall resistance to within a part per 10,000 and longitudinal resistivity under 1~$\Omega$ per square, with chiral edge transport explicitly confirmed by non-local measurements. Deviations from this behavior are found to be caused by thermally-activated carriers, which can be eliminated by taking advantage of an unexpected magnetocaloric effect.
\end{abstract}

\maketitle

The discovery of the quantum Hall effect (QHE)~\cite{KvK,Tsui} led to a new understanding of electronic behavior in which topology plays a central role~\cite{Laughlin,Thouless}. Initially, the critical experimental observation was the precise quantization of the Hall resistance to integer divisions of $h/e^2$, where $h$ is Planck's constant and $e$ is the electron charge. This quantization, immune to sample-specific disorder, now forms the basis for a metrological standard~\cite{Metrology}. A complementary feature\textemdash zero longitudinal resistance, reflecting resistanceless transport along sample edges\textemdash could also have technological applications, were it not for the demanding environmental requirements for achieving the QHE: a large magnetic field to break time-reversal symmetry (TRS) and, in most cases, cryogenic temperatures. Ideas for producing a similar phenomenology without an external magnetic field have long been considered~\cite{Haldane}, often involving the interplay of symmetry and topology in new material systems.

In the past decade, topological insulators (TIs) have emerged as a promising approach. In both two-dimensional~\cite{KaneMele,BernevigHughes,Molenkamp} and three-dimensional~\cite{FuKaneMele,QiHughes,Hasan,HasanReview,ZhangReview} forms, conduction in TIs is restricted to topologically-protected boundary states. In the 3D case, the presence of ferromagnetic exchange can break TRS, opening a gap in the otherwise Dirac-like surface states~\cite{YuQAHE,Nagaosa,Yulin}. But topology adds a twist: even a uniformly magnetized sample will have, relative to the normal vector of the surface, a domain boundary where the magnetization switches from inward to outward. Along this line the gap should close, restoring conduction~\cite{Nagaosa}. In a thin film geometry in which the easy axis of the magnetism is out-of-plane, confinement along the sample side wall should ensure conduction is one-dimensional while the surface gradient of the magnetism restricts it to only one direction, leading to ballistic, chiral transport. In a Hall bar geometry, this would be observed as the quantum anomalous Hall effect (QAHE), with a zero longitudinal resistance and a transverse resistance quantized to $h/ne^2$, where $n$ is typically $\pm$1 but can in principle be a higher integer given sufficiently strong exchange~\cite{JingPlateaus}.

Experimental realization of the QAHE has been swift. Doping films of the ternary TI family (Bi,Sb)$_{2}$Te$_{3}$ with Mn or Cr was found to produce robust out-of-plane ferromagnetism and a large anomalous Hall effect in transport~\cite{Tokurapre,Tsinghuapre,UCLApre}. Further growth optimization and chemical potential manipulation led to the recent achievement of the full quantized effect~\cite{Tsinghua,UCLA,Tokura}, albeit at dilution refrigerator temperatures. In two cases~\cite{Tsinghua,Tokura}, an applied magnetic field was necessary to decrease the longitudinal resistivity, presumably to eliminate other conduction channels. The possibilities for these channels include non-chiral edge modes~\cite{Jing}, variable-range hopping, or band transport of thermally populated 2D surface or 3D bulk carriers. In contrast, the device measured in Kou \emph{et al.}~\cite{UCLA} demonstrates its lowest longitudinal resistance near zero applied field.

\begin{figure}
\includegraphics[width=\columnwidth]{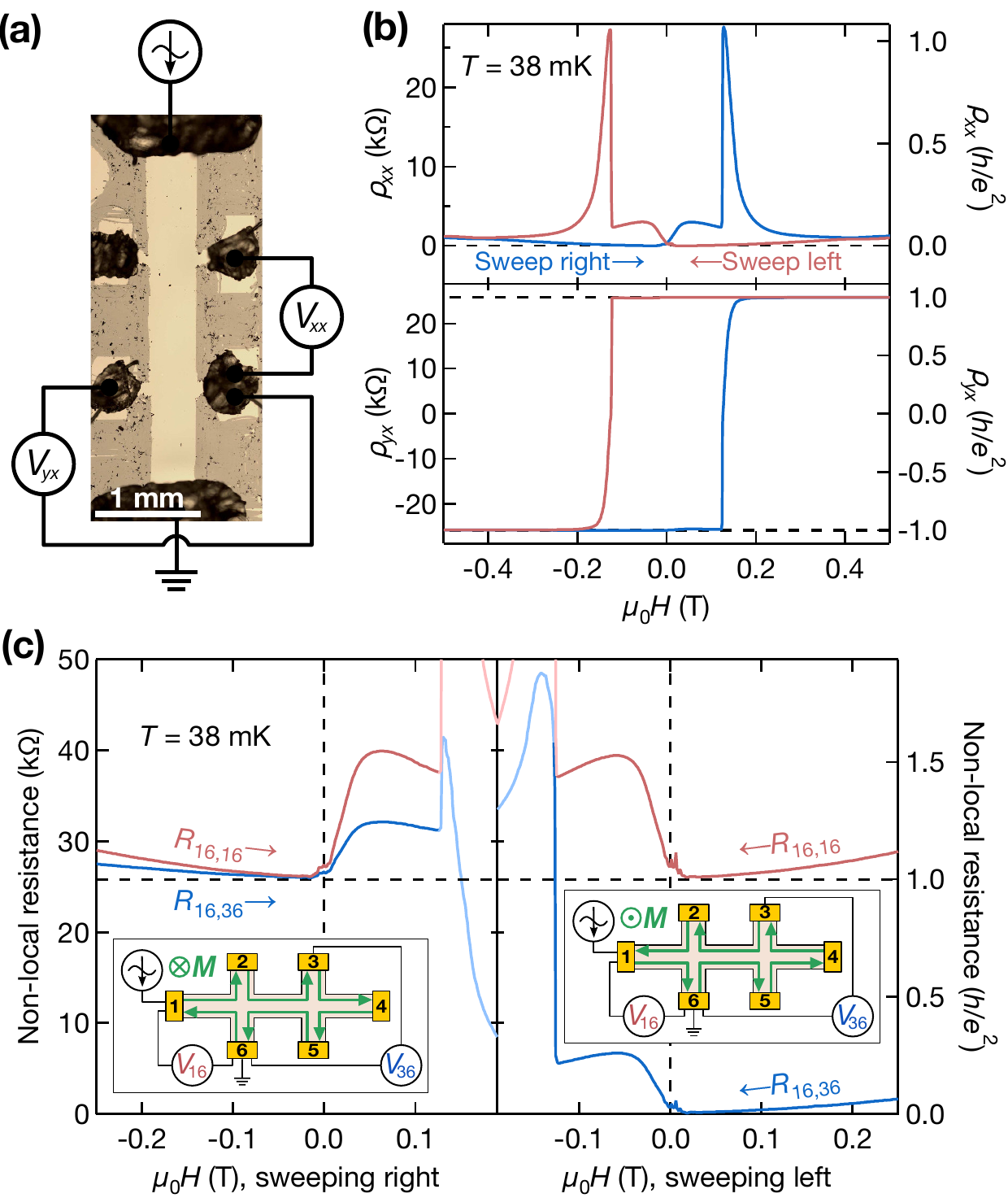}
\caption{\label{fig1} Device demonstrating quantum anomalous Hall effect. (a), Photograph of 10 nm-thick film of (Cr$_{0.12}$Bi$_{0.26}$Sb$_{0.62}$)$_{2}$Te$_{3}$ on a GaAs substrate, scratched by hand into Hall bar shape, with indium metal ohmic contacts. Schematic measurement setup included. (b), Longitudinal resistivity $\rho_{xx}$ and transverse resistivity $\rho_{yx}$ of the device at base temperature as a function of the applied magnetic field $\mu_{0}H$ in each sweep direction, forming a ferromagnetic hysteresis loop. As the field approaches zero from either starting point, $\rho_{yx}$ reaches its quantized value $h/e^2$ and $\rho_{xx}$ approaches zero. (c), Non-local measurements verifying edge-dominated transport. The insets show the measurements performed and chirality at each magnetization.}
\end{figure}

In this Letter, we study the QAHE in this regime using material from the same growth, demonstrating the hallmarks of the effect: vanishing longitudinal resistance, precisely quantized Hall resistance that switches sign with magnetization, and direct confirmation of edge transport, all in the absence of an applied field. Where deviations from this ideal behavior occur, we attribute them to thermally-activated carriers whose presence can be fortuitously controlled by demagnetization cooling of some other magnetic system in the sample. The material is 10 quintuple layers of (Cr$_{0.12}$Bi$_{0.26}$Sb$_{0.62}$)$_{2}$Te$_{3}$ grown via molecular beam epitaxy and capped with alumina on a semi-insulating GaAs substrate. To avoid possible doping of the film through lithographic processing, following~\cite{Tsinghua} we use a sharp metal tip to scratch the film into a Hall bar shape, and form ohmic contacts by placing indium metal onto each of the six terminals. The region between the voltage leads is 1.1 mm long and 0.45 mm wide (Fig.\ \ref{fig1}(a)). Four-terminal resistances are measured via standard lock-in amplifier techniques~\cite{SuppInfo} with the sample in a dilution refrigerator with its mixing chamber cooled to 38 mK. We calibrate the aggregate amplifier gain of the setup by measuring a conventional $\nu = 1$ quantum Hall plateau on a separate high-mobility graphene sample~\cite{SuppInfo}.

At base temperature we reproduce the ferromagnetic hysteresis loop measured by the anomalous Hall effect in Kou \emph{et al.}~\cite{UCLA} (Fig.\ \ref{fig1}(b)). The sign of the transverse (Hall) resistivity, $\rho_{yx}$, reflects the device's magnetization direction, $M_{z}$, which we can set to positive (``+1") or negative (``-1") by applying a field $\mu_{0}H$ with magnitude greater than the 125 mT coercive field. As we sweep $H$ toward zero, $\rho_{yx}$ reaches its quantized value $\pm h/e^2 \approx \pm$25,813 $\Omega$ while the longitudinal resistivity $\rho_{xx}$ decreases precipitously (in one case, reaching as low as 15 $\Omega$). After crossing zero field, $\rho_{xx}$ increases to a few k$\Omega$ before spiking higher at the coercive field as $\rho_{yx}$ changes sign. Both measurements settle toward their quantized values as $|H|$ increases, but only reach full quantization on the return arm of the hysteresis loop, again just before zero field.

Although the resistivity tensor takes on the expected values, the hysteresis loop does not directly verify that edge conduction dominates in this regime. Non-local measurement configurations, such as that shown schematically in Fig.\ \ref{fig1}(c), are one way to establish this~\cite{MolenkampHbars}. In the limit of chiral, ballistic edge transport, the chemical potential along the chirality direction only changes at leads that act as a current sources or drains, as prescribed by Landauer-B{\"u}ttiker formalism~\cite{Buttiker} and demonstrated in the QHE~\cite{Beenakker} (though not explicitly, to date, in the QAHE). For example, while flowing current between adjacent contacts (labelled 1 and 6), the remaining four contacts should maintain the same voltage as either the current source or drain, depending on whether the QAHE chirality is clockwise or counterclockwise, respectively. In Fig.\ \ref{fig1}(c) we measure the voltage drop from a contact on the opposite side of the device (3) to the drain (6), and plot the resulting three-terminal resistance $R_{16,36}$ compared to the two-terminal value $R_{16,16}$. At negative magnetization (left panel), corresponding to clockwise equilibration, both quantities approach the ballistic value $h/e^2$, indicating that contact 3 is nearly equilibrated with the source. At the opposite magnetization, where voltages are propagated counterclockwise (right panel), $R_{16,16}$ approaches $h/e^2$ while $R_{16,36}$ approaches zero due to the equilibration of contact 3 with the drain. Near zero field, the deviations from idealized behavior (in all cases under 200 $\Omega$), likely reflect contact resistances and possibly the presence of extra dissipative helical edge modes~\cite{Jing}. Using one of the remaining contacts in the role of contact 3 results in the same behavior~\cite{SuppInfo}.

\begin{figure}
\includegraphics[width=\columnwidth]{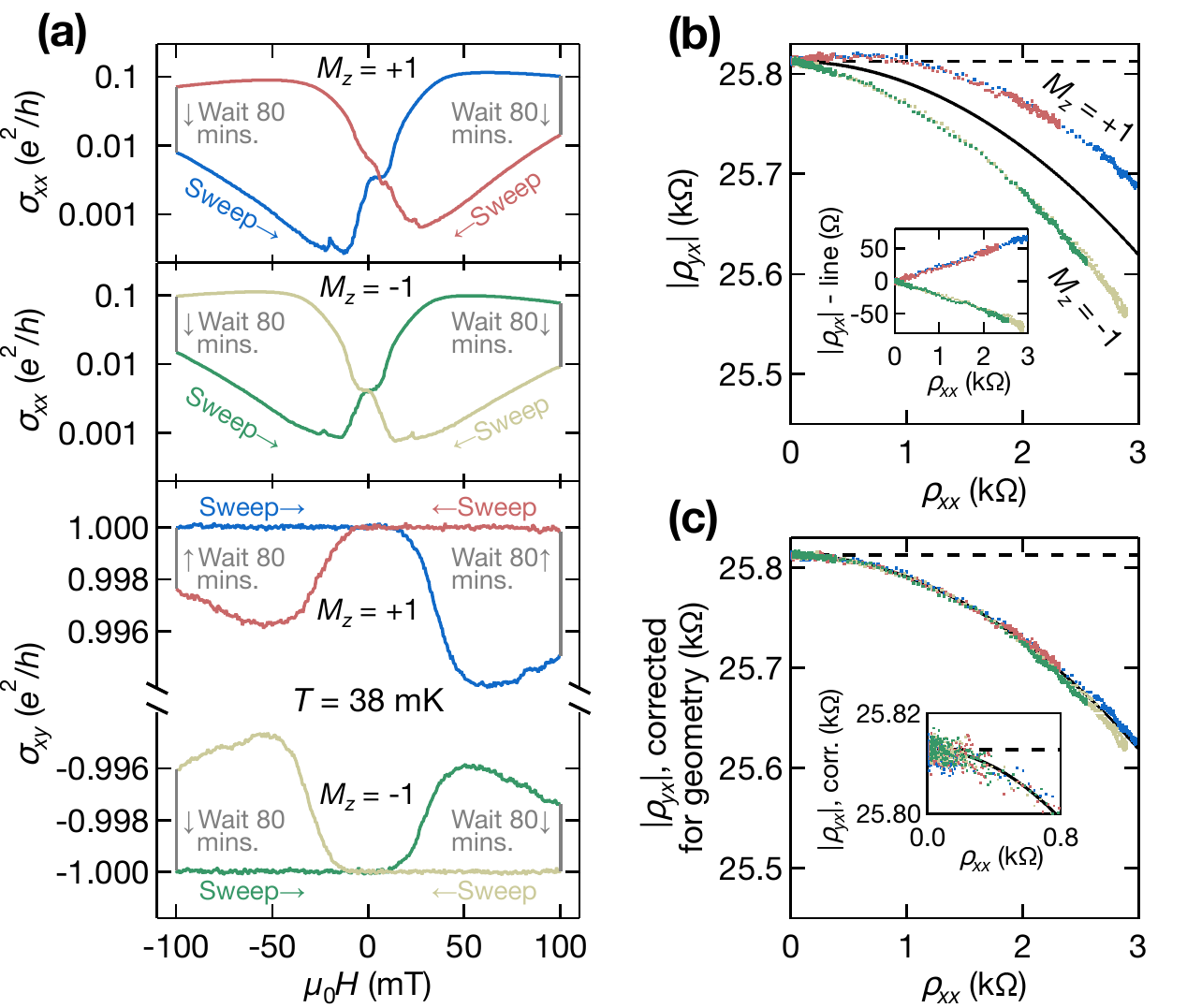}
\caption{\label{fig2} Precise quantization near zero applied field. (a), Longitudinal and transverse conductivities in hysteresis loops over field ranges smaller than the coercive field so as to maintain starting magnetization $M_{z}$. When approaching zero field from either starting point, $\sigma_{xy} = e^2/h$ to 0.01\% precision while $\sigma_{xx}$ reaches as low as 0.0002 $e^2/h$. (b), (c), Resistivities measured during the hysteresis loops, plotted parametrically, both before (b) and after (c) performing a correction for geometry. The inset of (b) shows the linear deviation of the two magnetization branches from a parabolic line, resulting from uneven spacing of the leads. The inset of (c) shows a close-up of the corrected resistivity data, with $\rho_{yx}$ quantized to $h/e^2$ within 3 $\Omega$ at any given point when $\rho_{xx} <$ 200 $\Omega$.}
\end{figure}

Returning to four-terminal measurements, we find that the best quantization can be obtained by maintaining the film's magnetization (i.e. keeping $|\mu_{0}H|$ smaller than the coercive field) but following the ``hysteresis loops" shown in Figure \ref{fig2}(a). Starting from any combination of magnetization and field polarity, sweeping $H$ toward zero suppresses the longitudinal conductivity $\sigma_{xx} = \rho_{xx}/(\rho_{xx}^2+\rho_{yx}^2)$ to as low as 0.0003 $e^2/h$ while the transverse conductivity $\sigma_{xy} = \rho_{yx}/(\rho_{xx}^2+\rho_{yx}^2)$ remains quantized to within 0.01\% of $e^2/h$. Passing through zero and then increasing $|H|$ destroys this quantization, which we can recover by waiting at constant field for 80 minutes and then sweeping back toward zero.

These reported conductivity values have undergone one correction for imperfect device geometry. Uneven spacing between the voltage probes of the Hall bar can add a small component of $\rho_{xx}$ to the measured value of $\rho_{yx}$. (In nonmagnetic samples, this is conventionally corrected by antisymmetrizing $\rho_{yx}$ about zero field.) In a parametric plot of the resistivity data from the same hysteresis loops (Fig.\ \ref{fig2}(b)), with $|\rho_{yx}|$ along the y-axis and $\rho_{xx}$ along the x-axis, we observe an asymmetry between the two magnetizations. Each branch deviates from a parabolic arc (the expected leading order contribution), with the size of the deviation explicitly verified to grow linearly with $\rho_{xx}$ with a coefficient of 2\% (Fig.\ \ref{fig2}(b), inset). By taking field sweeps at $T =$ 40 K, above the film's Curie temperature~\cite{UCLA}, we obtain an independent but matching measure of this geometric mixing coefficient, which is also verified by a numerical Poisson simulation of current flow~\cite{SuppInfo}. After removing the $\rho_{xx}$ component from $\rho_{yx}$ the data nearly collapse onto a single curve (Fig.\ \ref{fig2}(c)). In the vicinity of vanishing $\rho_{xx}$, the parametric plot after this correction demonstrates quantization of $\rho_{yx}$ to within $\pm$3 $\Omega$ (Fig.\ \ref{fig2}(c), inset).

\begin{figure}[b]
\includegraphics[width=\columnwidth]{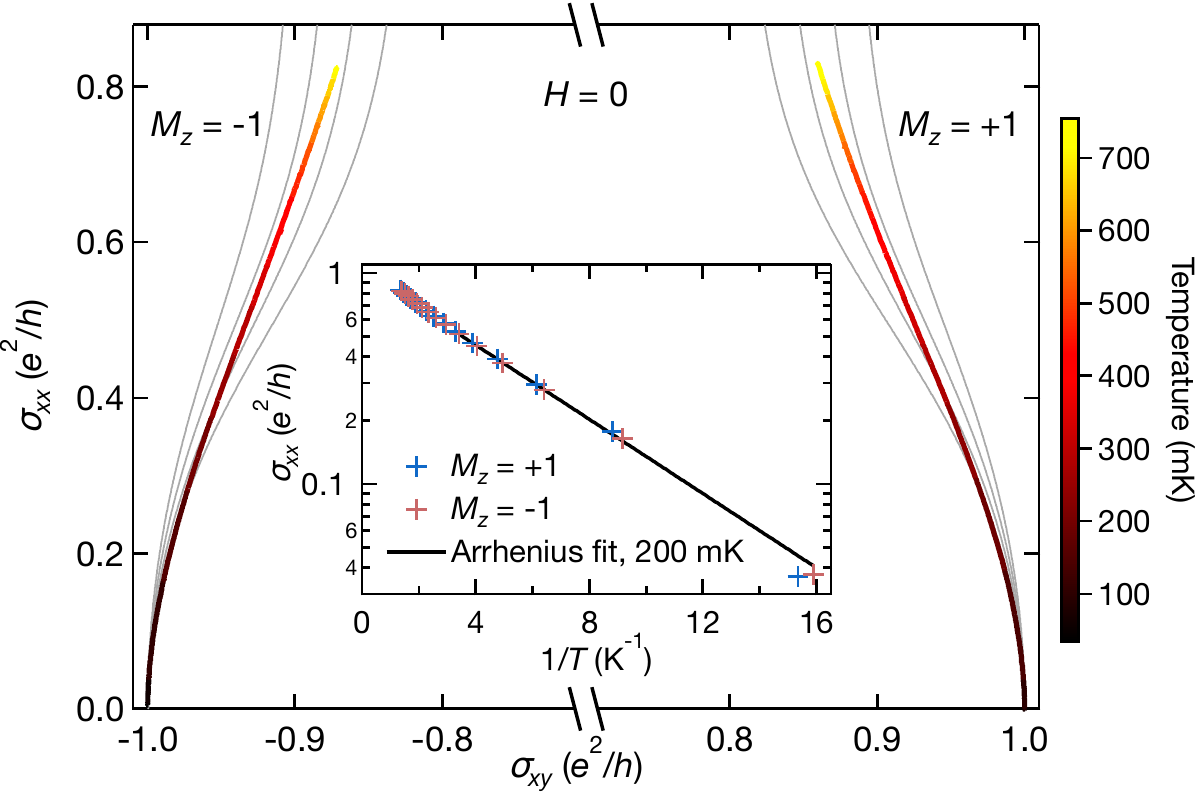}
\caption{\label{fig3} Temperature dependence implying possible thermal activation. Parametric plot of $\rho_{xx}$ versus $\rho_{yx}$ at each magnetization while increasing temperature, with calculated renormalization group flow lines shown in gray. Inset, an Arrhenius plot of $\sigma_{xx}$ as a function of inverse temperature, showing roughly exponential dependence. The implied energy scale of the fit (for instance, if conductivity is explained by thermal activation over a gap) is 17 $\mu$eV, or 200 mK.}
\end{figure}

The system stays on this curve in resistivity space, even during the wait times when $\rho_{xx}$ falls, suggesting that the position along the arc is determined by some parameter other than magnetization or applied field. Temperature is an obvious possibility. We extend the relationship by warming the sample above 100 mK, inducing a large longitudinal conductivity that becomes comparable to $e^2/h$ by 750 mK (Fig.\ \ref{fig3}). The trajectory of the QAHE in conductivity space during this process has been studied previously, and found to obey the same symmetry laws and renormalization group properties as the QHE~\cite{Tokura}. We find reasonable qualitative agreement with calculated~\cite{RGflowlines} renormalization group flow lines (Fig.\ \ref{fig3}, gray lines).

\begin{figure*}
\includegraphics[width=\hsize]{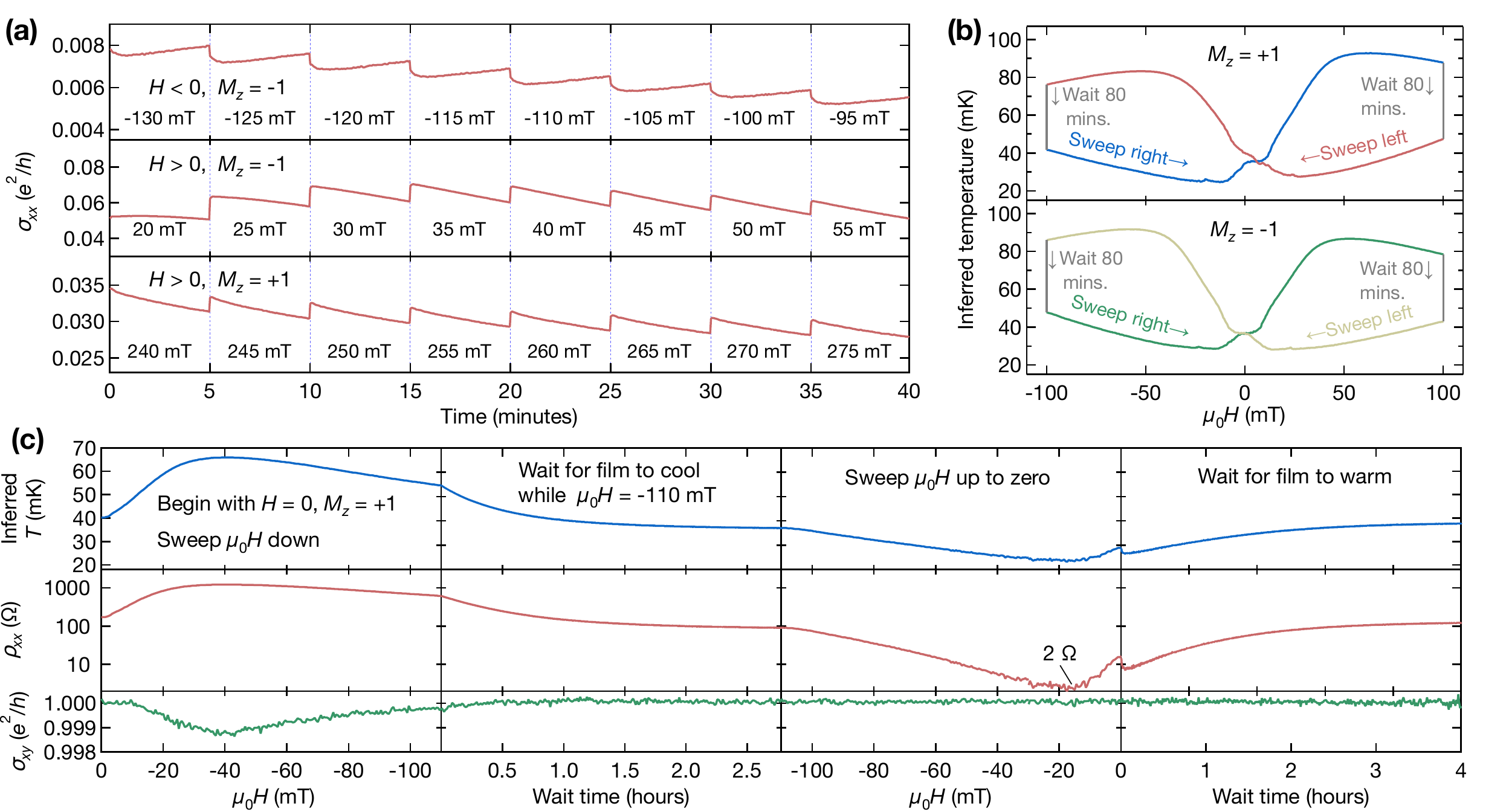}
\caption{\label{fig4} Evidence for demagnetization changing temperature. (a), Effect of upward field sweep on $\sigma_{xx}$ in three field/polarity regions, consistent with demagnetization hypothesis. When $H<0$ (decreasing $|H|$), $\sigma_{xx}$ decreases when the field changes and creeps back up during a five-minute wait. When $H>0$ (increasing $|H|$), the opposite occurs regardless of magnetization. (b), Inferred temperature of the TI film, extrapolated using the temperature fit in Figure \ref{fig3}, during the hysteresis loop in Figure \ref{fig2}(a). After a long wait, the temperature approaches the fridge thermometer reading, $\sim$40 mK, whereas sweeping $|H|$ down drives the temperature below 30 mK and sweeping $|H|$ up drives it above 90 mK. (c), A rudimentary demagnetization cycle, involving a slow field sweep to magnetize and a fast field sweep to zero to perform demagnetization cooling. Plotted during this cycle are the inferred temperature, $\rho_{xx}$ (now dropping below 2 $\Omega$), and $\sigma_{xy}$ which remains within 0.01\% of $e^2/h$ during the last three stages.}
\end{figure*}

To obtain a functional form of the temperature dependence, we plot $\sigma_{xx}$ versus reciprocal temperature (Fig.\ \ref{fig3}, inset). Such plots, useful for identifying conduction that is thermally activated over an energy barrier, are commonly used in quantum Hall systems to extract gap sizes~\cite{Tsui}. Here, an Arrhenius (exponential) fit holds down to 60 mK, suggesting that the nonzero $\sigma_{xx}$ represents carriers thermally activated into a surface band. We emphasize that this is not a direct measurement of the size of the exchange gap: with no gate electrode, the chemical potential cannot be tuned to the middle of the surface gap. Furthermore, the fit continues to work well above the extracted energy scale, where describing thermal activation requires use of the Fermi-Dirac distribution as well as knowledge of the system's density of states and mobility as a function of energy~\cite{SuppInfo}. Still, the data show some form of exponential activation suggesting that the nearest band edge is 17 $\mu$eV away, as indicated by a 200 mK characteristic temperature in the Arrhenius fit.

At the lowest temperatures, a sample's electron temperature can diverge from the cryostat's base temperature (here, the mixing chamber plate temperature). In fact, the wide variation of $\sigma_{xx}$ in Fig.\ \ref{fig2}(a) indicates that the temperature may be changing substantially during a hysteresis loop. We hypothesize that this takes place via magnetization or demagnetization of some spin system in the sample that heats or cools (respectively) the sample's electrons. We test this hypothesis by sweeping $\mu_{0}H$ from negative to positive, and stopping for five minutes every 5 mT to allow the sample temperature to partially equilibrate with the refrigerator. In a field region where $H$ is negative and its magnitude is decreasing (Fig.\ \ref{fig4}(a), top panel), $\sigma_{xx}$ decreases during the field step and then creeps back up during the wait time. We interpret this as demagnetization whose increase in entropy draws heat from the film, driving it to a lower temperature~\cite{MCE}, followed by a partial re-equilibration while we wait. At positive $H$ the opposite occurs (Fig.\ \ref{fig4}(a), middle panel). The pattern does not change after passing the coercive field and switching the film's magnetization (Fig.\ \ref{fig4}(a), bottom panel), indicating that only the polarity of the applied field matters.

As a further check on the reasonability of this hypothesis, we extrapolate the fit in Fig.\ \ref{fig3} down to lower temperatures and apply it to the conductance data in Fig.\ \ref{fig2}(d) to obtain the approximate sample temperature during a hysteresis loop. The result (Fig.\ \ref{fig4}(b)) implies that the electron temperature after an 80 minute wait at constant field approaches the mixing chamber thermometer reading, $\sim$40 mK. During downward field sweeps the sample drops below 30 mK, substantially below the refrigerator temperature, and during upward sweeps it exceeds 90 mK. Although the exact identity of the system responsible for this effect is unclear, a naive model of a paramagnet exchanging heat with the electrons qualitatively fits the shape of temperature variation during demagnetization and suggests a Land\'e g-factor of $\sim$0.15~\cite{SuppInfo}. We further believe it to be specific to and collocated with the TI sample: the refrigerator's thermometry shows minimal change during the hysteresis loop and, if we pause the loop during demagnetization, $\sigma_{xx}$ can remain below its equilibrium value for hours. Both details suggest that the TI surface and magnetic system are nearly thermally isolated from the cryostat when driven to their lowest temperatures~\cite{SuppInfo}.

If we wish to minimize $\rho_{xx}$ or improve quantization of $\sigma_{xy}$, an analogy to the traditional adiabatic demagnetization cycle is useful. A slow increase in the field's magnitude followed by a long wait, allowing maximal thermal equilibration along the way, approximates an isothermal magnetization step. Then a fast decrease in field magnitude (though not so fast that the heat load from the magnet sweep becomes relevant), produces adiabatic demagnetization. We plot $\rho_{xx}$, $\sigma_{xy}$, and the inferred temperature during this process in Figure \ref{fig4}(c). Indeed, we drive $\rho_{xx}$ to as low as 2 $\Omega$ ($\sigma_{xx} <$ 0.0001 $e^2/h$) and the temperature to 25 mK, with excellent quantization in $\sigma_{xy}$ along the way. We also include a long wait at the end, demonstrating the long timescale for reequilibration of $\rho_{xx}$. Modifying this process to end at a small nonzero applied field, we can even drive $\rho_{xx}$ below 1 $\Omega$~\cite{SuppInfo}. In future experiments, adding a gate electrode to optimize the position of the chemical potential in the gap may yield a completely vanishing longitudinal resistance.\\

\begin{acknowledgments}
Sample preparation, measurements, and analysis were supported by the U.S.\ Department of Energy, Office of Science, Basic Energy Sciences, under Award \#19-7503. Materials growth, surface characterization, preliminary electrical characterization, and electronic instrumentation were supported by the DARPA MESO program under Contracts No.\ N66001-12-1-4034 and No.\ N66001-11-1-4105. Infrastructure and cryostat support were funded in part by the Gordon and Betty Moore Foundation through Grant GBMF3429 to D.~G.~G. The authors acknowledge Stefan Grauer and Charles Gould for proposing the non-local measurements and for insightful technical discussions, Jing Wang, Biao Lian, and Xiaoliang Qi for theoretical comments, Yang Feng and Ke He for stimulating discussions about material properties, Jingsong Zhang for help with defining Hall bars, and Lucas Peeters for help with device measurement. K.~L.~W. acknowledges the support of the Raytheon endorsement. A.~J.~B. acknowledges support from a Benchmark Stanford Graduate Fellowship, E.~J.~F. acknowledges support from a DOE Office of Science Graduate Fellowship, and X.~K. acknowledges partial support from a Qualcomm Innovation Fellowship.\\
\end{acknowledgments}

\textit{Note added.}\textemdash During preparation of this Letter, we became aware of work by Chang \emph{et al.} that reports a similar degree of quantization in V-doped BiSbTe$_3$~\cite{MosesChan}.

\bibliography{QAHE}

\begin{thebibliography}{32}%
\makeatletter
\providecommand \@ifxundefined [1]{%
 \@ifx{#1\undefined}
}%
\providecommand \@ifnum [1]{%
 \ifnum #1\expandafter \@firstoftwo
 \else \expandafter \@secondoftwo
 \fi
}%
\providecommand \@ifx [1]{%
 \ifx #1\expandafter \@firstoftwo
 \else \expandafter \@secondoftwo
 \fi
}%
\providecommand \natexlab [1]{#1}%
\providecommand \enquote  [1]{``#1''}%
\providecommand \bibnamefont  [1]{#1}%
\providecommand \bibfnamefont [1]{#1}%
\providecommand \citenamefont [1]{#1}%
\providecommand \href@noop [0]{\@secondoftwo}%
\providecommand \href [0]{\begingroup \@sanitize@url \@href}%
\providecommand \@href[1]{\@@startlink{#1}\@@href}%
\providecommand \@@href[1]{\endgroup#1\@@endlink}%
\providecommand \@sanitize@url [0]{\catcode `\\12\catcode `\$12\catcode
  `\&12\catcode `\#12\catcode `\^12\catcode `\_12\catcode `\%12\relax}%
\providecommand \@@startlink[1]{}%
\providecommand \@@endlink[0]{}%
\providecommand \url  [0]{\begingroup\@sanitize@url \@url }%
\providecommand \@url [1]{\endgroup\@href {#1}{\urlprefix }}%
\providecommand \urlprefix  [0]{URL }%
\providecommand \Eprint [0]{\href }%
\providecommand \doibase [0]{http://dx.doi.org/}%
\providecommand \selectlanguage [0]{\@gobble}%
\providecommand \bibinfo  [0]{\@secondoftwo}%
\providecommand \bibfield  [0]{\@secondoftwo}%
\providecommand \translation [1]{[#1]}%
\providecommand \BibitemOpen [0]{}%
\providecommand \bibitemStop [0]{}%
\providecommand \bibitemNoStop [0]{.\EOS\space}%
\providecommand \EOS [0]{\spacefactor3000\relax}%
\providecommand \BibitemShut  [1]{\csname bibitem#1\endcsname}%
\let\auto@bib@innerbib\@empty
\bibitem [{\citenamefont {Klitzing}\ \emph {et~al.}(1980)\citenamefont
  {Klitzing}, \citenamefont {Dorda},\ and\ \citenamefont {Pepper}}]{KvK}%
  \BibitemOpen
  \bibfield  {author} {\bibinfo {author} {\bibfnamefont {K.~v.}\ \bibnamefont
  {Klitzing}}, \bibinfo {author} {\bibfnamefont {G.}~\bibnamefont {Dorda}}, \
  and\ \bibinfo {author} {\bibfnamefont {M.}~\bibnamefont {Pepper}},\ }\href
  {\doibase 10.1103/PhysRevLett.45.494} {\bibfield  {journal} {\bibinfo
  {journal} {Phys. Rev. Lett.}\ }\textbf {\bibinfo {volume} {45}},\ \bibinfo
  {pages} {494} (\bibinfo {year} {1980})}\BibitemShut {NoStop}%
\bibitem [{\citenamefont {Tsui}\ \emph {et~al.}(1982)\citenamefont {Tsui},
  \citenamefont {Stormer},\ and\ \citenamefont {Gossard}}]{Tsui}%
  \BibitemOpen
  \bibfield  {author} {\bibinfo {author} {\bibfnamefont {D.~C.}\ \bibnamefont
  {Tsui}}, \bibinfo {author} {\bibfnamefont {H.~L.}\ \bibnamefont {Stormer}}, \
  and\ \bibinfo {author} {\bibfnamefont {A.~C.}\ \bibnamefont {Gossard}},\
  }\href {\doibase 10.1103/PhysRevLett.48.1559} {\bibfield  {journal} {\bibinfo
   {journal} {Phys. Rev. Lett.}\ }\textbf {\bibinfo {volume} {48}},\ \bibinfo
  {pages} {1559} (\bibinfo {year} {1982})}\BibitemShut {NoStop}%
\bibitem [{\citenamefont {Laughlin}(1981)}]{Laughlin}%
  \BibitemOpen
  \bibfield  {author} {\bibinfo {author} {\bibfnamefont {R.~B.}\ \bibnamefont
  {Laughlin}},\ }\href {\doibase 10.1103/PhysRevB.23.5632} {\bibfield
  {journal} {\bibinfo  {journal} {Phys. Rev. B}\ }\textbf {\bibinfo {volume}
  {23}},\ \bibinfo {pages} {5632} (\bibinfo {year} {1981})}\BibitemShut
  {NoStop}%
\bibitem [{\citenamefont {Thouless}\ \emph {et~al.}(1982)\citenamefont
  {Thouless}, \citenamefont {Kohmoto}, \citenamefont {Nightingale},\ and\
  \citenamefont {den Nijs}}]{Thouless}%
  \BibitemOpen
  \bibfield  {author} {\bibinfo {author} {\bibfnamefont {D.~J.}\ \bibnamefont
  {Thouless}}, \bibinfo {author} {\bibfnamefont {M.}~\bibnamefont {Kohmoto}},
  \bibinfo {author} {\bibfnamefont {M.~P.}\ \bibnamefont {Nightingale}}, \ and\
  \bibinfo {author} {\bibfnamefont {M.}~\bibnamefont {den Nijs}},\ }\href
  {\doibase 10.1103/PhysRevLett.49.405} {\bibfield  {journal} {\bibinfo
  {journal} {Phys. Rev. Lett.}\ }\textbf {\bibinfo {volume} {49}},\ \bibinfo
  {pages} {405} (\bibinfo {year} {1982})}\BibitemShut {NoStop}%
\bibitem [{\citenamefont {von Klitzing}(2005)}]{Metrology}%
  \BibitemOpen
  \bibfield  {author} {\bibinfo {author} {\bibfnamefont {K.}~\bibnamefont {von
  Klitzing}},\ }\href {\doibase 10.1098/rsta.2005.1640} {\bibfield  {journal}
  {\bibinfo  {journal} {Phil. Trans. R. Soc. A}\ }\textbf {\bibinfo {volume}
  {363}},\ \bibinfo {pages} {2203} (\bibinfo {year} {2005})}\BibitemShut
  {NoStop}%
\bibitem [{\citenamefont {Haldane}(1988)}]{Haldane}%
  \BibitemOpen
  \bibfield  {author} {\bibinfo {author} {\bibfnamefont {F.~D.~M.}\
  \bibnamefont {Haldane}},\ }\href {\doibase 10.1103/PhysRevLett.61.2015}
  {\bibfield  {journal} {\bibinfo  {journal} {Phys. Rev. Lett.}\ }\textbf
  {\bibinfo {volume} {61}},\ \bibinfo {pages} {2015} (\bibinfo {year}
  {1988})}\BibitemShut {NoStop}%
\bibitem [{\citenamefont {Kane}\ and\ \citenamefont {Mele}(2005)}]{KaneMele}%
  \BibitemOpen
  \bibfield  {author} {\bibinfo {author} {\bibfnamefont {C.~L.}\ \bibnamefont
  {Kane}}\ and\ \bibinfo {author} {\bibfnamefont {E.~J.}\ \bibnamefont
  {Mele}},\ }\href {\doibase 10.1103/PhysRevLett.95.146802} {\bibfield
  {journal} {\bibinfo  {journal} {Phys. Rev. Lett.}\ }\textbf {\bibinfo
  {volume} {95}},\ \bibinfo {pages} {146802} (\bibinfo {year}
  {2005})}\BibitemShut {NoStop}%
\bibitem [{\citenamefont {Bernevig}\ \emph {et~al.}(2006)\citenamefont
  {Bernevig}, \citenamefont {Hughes},\ and\ \citenamefont
  {Zhang}}]{BernevigHughes}%
  \BibitemOpen
  \bibfield  {author} {\bibinfo {author} {\bibfnamefont {B.~A.}\ \bibnamefont
  {Bernevig}}, \bibinfo {author} {\bibfnamefont {T.~L.}\ \bibnamefont
  {Hughes}}, \ and\ \bibinfo {author} {\bibfnamefont {S.-C.}\ \bibnamefont
  {Zhang}},\ }\href {\doibase 10.1126/science.1133734} {\bibfield  {journal}
  {\bibinfo  {journal} {Science}\ }\textbf {\bibinfo {volume} {314}},\ \bibinfo
  {pages} {1757} (\bibinfo {year} {2006})}\BibitemShut {NoStop}%
\bibitem [{\citenamefont {K\"onig}\ \emph {et~al.}(2007)\citenamefont
  {K\"onig}, \citenamefont {Wiedmann}, \citenamefont {Br\"une}, \citenamefont
  {Roth}, \citenamefont {Buhmann}, \citenamefont {Molenkamp}, \citenamefont
  {Qi},\ and\ \citenamefont {Zhang}}]{Molenkamp}%
  \BibitemOpen
  \bibfield  {author} {\bibinfo {author} {\bibfnamefont {M.}~\bibnamefont
  {K\"onig}}, \bibinfo {author} {\bibfnamefont {S.}~\bibnamefont {Wiedmann}},
  \bibinfo {author} {\bibfnamefont {C.}~\bibnamefont {Br\"une}}, \bibinfo
  {author} {\bibfnamefont {A.}~\bibnamefont {Roth}}, \bibinfo {author}
  {\bibfnamefont {H.}~\bibnamefont {Buhmann}}, \bibinfo {author} {\bibfnamefont
  {L.~W.}\ \bibnamefont {Molenkamp}}, \bibinfo {author} {\bibfnamefont {X.-L.}\
  \bibnamefont {Qi}}, \ and\ \bibinfo {author} {\bibfnamefont {S.-C.}\
  \bibnamefont {Zhang}},\ }\href {\doibase 10.1126/science.1148047} {\bibfield
  {journal} {\bibinfo  {journal} {Science}\ }\textbf {\bibinfo {volume}
  {318}},\ \bibinfo {pages} {766} (\bibinfo {year} {2007})}\BibitemShut
  {NoStop}%
\bibitem [{\citenamefont {Fu}\ \emph {et~al.}(2007)\citenamefont {Fu},
  \citenamefont {Kane},\ and\ \citenamefont {Mele}}]{FuKaneMele}%
  \BibitemOpen
  \bibfield  {author} {\bibinfo {author} {\bibfnamefont {L.}~\bibnamefont
  {Fu}}, \bibinfo {author} {\bibfnamefont {C.~L.}\ \bibnamefont {Kane}}, \ and\
  \bibinfo {author} {\bibfnamefont {E.~J.}\ \bibnamefont {Mele}},\ }\href
  {\doibase 10.1103/PhysRevLett.98.106803} {\bibfield  {journal} {\bibinfo
  {journal} {Phys. Rev. Lett.}\ }\textbf {\bibinfo {volume} {98}},\ \bibinfo
  {pages} {106803} (\bibinfo {year} {2007})}\BibitemShut {NoStop}%
\bibitem [{\citenamefont {Qi}\ \emph {et~al.}(2008)\citenamefont {Qi},
  \citenamefont {Hughes},\ and\ \citenamefont {Zhang}}]{QiHughes}%
  \BibitemOpen
  \bibfield  {author} {\bibinfo {author} {\bibfnamefont {X.-L.}\ \bibnamefont
  {Qi}}, \bibinfo {author} {\bibfnamefont {T.~L.}\ \bibnamefont {Hughes}}, \
  and\ \bibinfo {author} {\bibfnamefont {S.-C.}\ \bibnamefont {Zhang}},\ }\href
  {\doibase 10.1103/PhysRevB.78.195424} {\bibfield  {journal} {\bibinfo
  {journal} {Phys. Rev. B}\ }\textbf {\bibinfo {volume} {78}},\ \bibinfo
  {pages} {195424} (\bibinfo {year} {2008})}\BibitemShut {NoStop}%
\bibitem [{\citenamefont {Hsieh}\ \emph {et~al.}(2009)\citenamefont {Hsieh},
  \citenamefont {Xia}, \citenamefont {Qian}, \citenamefont {Wray},
  \citenamefont {Dil}, \citenamefont {Meier}, \citenamefont {Osterwalder},
  \citenamefont {Patthey}, \citenamefont {Checkelsky}, \citenamefont {Ong},
  \citenamefont {Fedorov}, \citenamefont {Lin}, \citenamefont {Bansil},
  \citenamefont {Grauer}, \citenamefont {Hor}, \citenamefont {Cava},\ and\
  \citenamefont {Hasan}}]{Hasan}%
  \BibitemOpen
  \bibfield  {author} {\bibinfo {author} {\bibfnamefont {D.}~\bibnamefont
  {Hsieh}}, \bibinfo {author} {\bibfnamefont {Y.}~\bibnamefont {Xia}}, \bibinfo
  {author} {\bibfnamefont {D.}~\bibnamefont {Qian}}, \bibinfo {author}
  {\bibfnamefont {L.}~\bibnamefont {Wray}}, \bibinfo {author} {\bibfnamefont
  {J.~H.}\ \bibnamefont {Dil}}, \bibinfo {author} {\bibfnamefont
  {F.}~\bibnamefont {Meier}}, \bibinfo {author} {\bibfnamefont
  {J.}~\bibnamefont {Osterwalder}}, \bibinfo {author} {\bibfnamefont
  {L.}~\bibnamefont {Patthey}}, \bibinfo {author} {\bibfnamefont {J.~G.}\
  \bibnamefont {Checkelsky}}, \bibinfo {author} {\bibfnamefont {N.~P.}\
  \bibnamefont {Ong}}, \bibinfo {author} {\bibfnamefont {A.~V.}\ \bibnamefont
  {Fedorov}}, \bibinfo {author} {\bibfnamefont {H.}~\bibnamefont {Lin}},
  \bibinfo {author} {\bibfnamefont {A.}~\bibnamefont {Bansil}}, \bibinfo
  {author} {\bibfnamefont {D.}~\bibnamefont {Grauer}}, \bibinfo {author}
  {\bibfnamefont {Y.~S.}\ \bibnamefont {Hor}}, \bibinfo {author} {\bibfnamefont
  {R.~J.}\ \bibnamefont {Cava}}, \ and\ \bibinfo {author} {\bibfnamefont
  {M.~Z.}\ \bibnamefont {Hasan}},\ }\href
  {http://dx.doi.org/10.1038/nature08234} {\bibfield  {journal} {\bibinfo
  {journal} {Nature}\ }\textbf {\bibinfo {volume} {460}},\ \bibinfo {pages}
  {1101} (\bibinfo {year} {2009})}\BibitemShut {NoStop}%
\bibitem [{\citenamefont {Hasan}\ and\ \citenamefont
  {Kane}(2010)}]{HasanReview}%
  \BibitemOpen
  \bibfield  {author} {\bibinfo {author} {\bibfnamefont {M.~Z.}\ \bibnamefont
  {Hasan}}\ and\ \bibinfo {author} {\bibfnamefont {C.~L.}\ \bibnamefont
  {Kane}},\ }\href {\doibase 10.1103/RevModPhys.82.3045} {\bibfield  {journal}
  {\bibinfo  {journal} {Rev. Mod. Phys.}\ }\textbf {\bibinfo {volume} {82}},\
  \bibinfo {pages} {3045} (\bibinfo {year} {2010})}\BibitemShut {NoStop}%
\bibitem [{\citenamefont {Qi}\ and\ \citenamefont {Zhang}(2011)}]{ZhangReview}%
  \BibitemOpen
  \bibfield  {author} {\bibinfo {author} {\bibfnamefont {X.-L.}\ \bibnamefont
  {Qi}}\ and\ \bibinfo {author} {\bibfnamefont {S.-C.}\ \bibnamefont {Zhang}},\
  }\href {\doibase 10.1103/RevModPhys.83.1057} {\bibfield  {journal} {\bibinfo
  {journal} {Rev. Mod. Phys.}\ }\textbf {\bibinfo {volume} {83}},\ \bibinfo
  {pages} {1057} (\bibinfo {year} {2011})}\BibitemShut {NoStop}%
\bibitem [{\citenamefont {Yu}\ \emph {et~al.}(2010)\citenamefont {Yu},
  \citenamefont {Zhang}, \citenamefont {Zhang}, \citenamefont {Zhang},
  \citenamefont {Dai},\ and\ \citenamefont {Fang}}]{YuQAHE}%
  \BibitemOpen
  \bibfield  {author} {\bibinfo {author} {\bibfnamefont {R.}~\bibnamefont
  {Yu}}, \bibinfo {author} {\bibfnamefont {W.}~\bibnamefont {Zhang}}, \bibinfo
  {author} {\bibfnamefont {H.-J.}\ \bibnamefont {Zhang}}, \bibinfo {author}
  {\bibfnamefont {S.-C.}\ \bibnamefont {Zhang}}, \bibinfo {author}
  {\bibfnamefont {X.}~\bibnamefont {Dai}}, \ and\ \bibinfo {author}
  {\bibfnamefont {Z.}~\bibnamefont {Fang}},\ }\href {\doibase
  10.1126/science.1187485} {\bibfield  {journal} {\bibinfo  {journal}
  {Science}\ }\textbf {\bibinfo {volume} {329}},\ \bibinfo {pages} {61}
  (\bibinfo {year} {2010})}\BibitemShut {NoStop}%
\bibitem [{\citenamefont {Nomura}\ and\ \citenamefont
  {Nagaosa}(2011)}]{Nagaosa}%
  \BibitemOpen
  \bibfield  {author} {\bibinfo {author} {\bibfnamefont {K.}~\bibnamefont
  {Nomura}}\ and\ \bibinfo {author} {\bibfnamefont {N.}~\bibnamefont
  {Nagaosa}},\ }\href {\doibase 10.1103/PhysRevLett.106.166802} {\bibfield
  {journal} {\bibinfo  {journal} {Phys. Rev. Lett.}\ }\textbf {\bibinfo
  {volume} {106}},\ \bibinfo {pages} {166802} (\bibinfo {year}
  {2011})}\BibitemShut {NoStop}%
\bibitem [{\citenamefont {Chen}\ \emph {et~al.}(2010)\citenamefont {Chen},
  \citenamefont {Chu}, \citenamefont {Analytis}, \citenamefont {Liu},
  \citenamefont {Igarashi}, \citenamefont {Kuo}, \citenamefont {Qi},
  \citenamefont {Mo}, \citenamefont {Moore}, \citenamefont {Lu}, \citenamefont
  {Hashimoto}, \citenamefont {Sasagawa}, \citenamefont {Zhang}, \citenamefont
  {Fisher}, \citenamefont {Hussain},\ and\ \citenamefont {Shen}}]{Yulin}%
  \BibitemOpen
  \bibfield  {author} {\bibinfo {author} {\bibfnamefont {Y.~L.}\ \bibnamefont
  {Chen}}, \bibinfo {author} {\bibfnamefont {J.-H.}\ \bibnamefont {Chu}},
  \bibinfo {author} {\bibfnamefont {J.~G.}\ \bibnamefont {Analytis}}, \bibinfo
  {author} {\bibfnamefont {Z.~K.}\ \bibnamefont {Liu}}, \bibinfo {author}
  {\bibfnamefont {K.}~\bibnamefont {Igarashi}}, \bibinfo {author}
  {\bibfnamefont {H.-H.}\ \bibnamefont {Kuo}}, \bibinfo {author} {\bibfnamefont
  {X.~L.}\ \bibnamefont {Qi}}, \bibinfo {author} {\bibfnamefont {S.~K.}\
  \bibnamefont {Mo}}, \bibinfo {author} {\bibfnamefont {R.~G.}\ \bibnamefont
  {Moore}}, \bibinfo {author} {\bibfnamefont {D.~H.}\ \bibnamefont {Lu}},
  \bibinfo {author} {\bibfnamefont {M.}~\bibnamefont {Hashimoto}}, \bibinfo
  {author} {\bibfnamefont {T.}~\bibnamefont {Sasagawa}}, \bibinfo {author}
  {\bibfnamefont {S.~C.}\ \bibnamefont {Zhang}}, \bibinfo {author}
  {\bibfnamefont {I.~R.}\ \bibnamefont {Fisher}}, \bibinfo {author}
  {\bibfnamefont {Z.}~\bibnamefont {Hussain}}, \ and\ \bibinfo {author}
  {\bibfnamefont {Z.~X.}\ \bibnamefont {Shen}},\ }\href {\doibase
  10.1126/science.1189924} {\bibfield  {journal} {\bibinfo  {journal}
  {Science}\ }\textbf {\bibinfo {volume} {329}},\ \bibinfo {pages} {659}
  (\bibinfo {year} {2010})}\BibitemShut {NoStop}%
\bibitem [{\citenamefont {Wang}\ \emph
  {et~al.}(2013{\natexlab{a}})\citenamefont {Wang}, \citenamefont {Lian},
  \citenamefont {Zhang}, \citenamefont {Xu},\ and\ \citenamefont
  {Zhang}}]{JingPlateaus}%
  \BibitemOpen
  \bibfield  {author} {\bibinfo {author} {\bibfnamefont {J.}~\bibnamefont
  {Wang}}, \bibinfo {author} {\bibfnamefont {B.}~\bibnamefont {Lian}}, \bibinfo
  {author} {\bibfnamefont {H.}~\bibnamefont {Zhang}}, \bibinfo {author}
  {\bibfnamefont {Y.}~\bibnamefont {Xu}}, \ and\ \bibinfo {author}
  {\bibfnamefont {S.-C.}\ \bibnamefont {Zhang}},\ }\href {\doibase
  10.1103/PhysRevLett.111.136801} {\bibfield  {journal} {\bibinfo  {journal}
  {Phys. Rev. Lett.}\ }\textbf {\bibinfo {volume} {111}},\ \bibinfo {pages}
  {136801} (\bibinfo {year} {2013}{\natexlab{a}})}\BibitemShut {NoStop}%
\bibitem [{\citenamefont {Checkelsky}\ \emph {et~al.}(2012)\citenamefont
  {Checkelsky}, \citenamefont {Ye}, \citenamefont {Onose}, \citenamefont
  {Iwasa},\ and\ \citenamefont {Tokura}}]{Tokurapre}%
  \BibitemOpen
  \bibfield  {author} {\bibinfo {author} {\bibfnamefont {J.~G.}\ \bibnamefont
  {Checkelsky}}, \bibinfo {author} {\bibfnamefont {J.}~\bibnamefont {Ye}},
  \bibinfo {author} {\bibfnamefont {Y.}~\bibnamefont {Onose}}, \bibinfo
  {author} {\bibfnamefont {Y.}~\bibnamefont {Iwasa}}, \ and\ \bibinfo {author}
  {\bibfnamefont {Y.}~\bibnamefont {Tokura}},\ }\href
  {http://dx.doi.org/10.1038/nphys2388} {\bibfield  {journal} {\bibinfo
  {journal} {Nat. Phys.}\ }\textbf {\bibinfo {volume} {8}},\ \bibinfo {pages}
  {729} (\bibinfo {year} {2012})}\BibitemShut {NoStop}%
\bibitem [{\citenamefont {Chang}\ \emph
  {et~al.}(2013{\natexlab{a}})\citenamefont {Chang}, \citenamefont {Zhang},
  \citenamefont {Liu}, \citenamefont {Zhang}, \citenamefont {Feng},
  \citenamefont {Li}, \citenamefont {Wang}, \citenamefont {Chen}, \citenamefont
  {Dai}, \citenamefont {Fang}, \citenamefont {Qi}, \citenamefont {Zhang},
  \citenamefont {Wang}, \citenamefont {He}, \citenamefont {Ma},\ and\
  \citenamefont {Xue}}]{Tsinghuapre}%
  \BibitemOpen
  \bibfield  {author} {\bibinfo {author} {\bibfnamefont {C.-Z.}\ \bibnamefont
  {Chang}}, \bibinfo {author} {\bibfnamefont {J.}~\bibnamefont {Zhang}},
  \bibinfo {author} {\bibfnamefont {M.}~\bibnamefont {Liu}}, \bibinfo {author}
  {\bibfnamefont {Z.}~\bibnamefont {Zhang}}, \bibinfo {author} {\bibfnamefont
  {X.}~\bibnamefont {Feng}}, \bibinfo {author} {\bibfnamefont {K.}~\bibnamefont
  {Li}}, \bibinfo {author} {\bibfnamefont {L.-L.}\ \bibnamefont {Wang}},
  \bibinfo {author} {\bibfnamefont {X.}~\bibnamefont {Chen}}, \bibinfo {author}
  {\bibfnamefont {X.}~\bibnamefont {Dai}}, \bibinfo {author} {\bibfnamefont
  {Z.}~\bibnamefont {Fang}}, \bibinfo {author} {\bibfnamefont {X.-L.}\
  \bibnamefont {Qi}}, \bibinfo {author} {\bibfnamefont {S.-C.}\ \bibnamefont
  {Zhang}}, \bibinfo {author} {\bibfnamefont {Y.}~\bibnamefont {Wang}},
  \bibinfo {author} {\bibfnamefont {K.}~\bibnamefont {He}}, \bibinfo {author}
  {\bibfnamefont {X.-C.}\ \bibnamefont {Ma}}, \ and\ \bibinfo {author}
  {\bibfnamefont {Q.-K.}\ \bibnamefont {Xue}},\ }\href {\doibase
  10.1002/adma.201203493} {\bibfield  {journal} {\bibinfo  {journal} {Adv.
  Mater.}\ }\textbf {\bibinfo {volume} {25}},\ \bibinfo {pages} {1065}
  (\bibinfo {year} {2013}{\natexlab{a}})}\BibitemShut {NoStop}%
\bibitem [{\citenamefont {Kou}\ \emph {et~al.}(2013)\citenamefont {Kou},
  \citenamefont {Lang}, \citenamefont {Fan}, \citenamefont {Jiang},
  \citenamefont {Nie}, \citenamefont {Zhang}, \citenamefont {Jiang},
  \citenamefont {Wang}, \citenamefont {Yao}, \citenamefont {He},\ and\
  \citenamefont {Wang}}]{UCLApre}%
  \BibitemOpen
  \bibfield  {author} {\bibinfo {author} {\bibfnamefont {X.}~\bibnamefont
  {Kou}}, \bibinfo {author} {\bibfnamefont {M.}~\bibnamefont {Lang}}, \bibinfo
  {author} {\bibfnamefont {Y.}~\bibnamefont {Fan}}, \bibinfo {author}
  {\bibfnamefont {Y.}~\bibnamefont {Jiang}}, \bibinfo {author} {\bibfnamefont
  {T.}~\bibnamefont {Nie}}, \bibinfo {author} {\bibfnamefont {J.}~\bibnamefont
  {Zhang}}, \bibinfo {author} {\bibfnamefont {W.}~\bibnamefont {Jiang}},
  \bibinfo {author} {\bibfnamefont {Y.}~\bibnamefont {Wang}}, \bibinfo {author}
  {\bibfnamefont {Y.}~\bibnamefont {Yao}}, \bibinfo {author} {\bibfnamefont
  {L.}~\bibnamefont {He}}, \ and\ \bibinfo {author} {\bibfnamefont {K.~L.}\
  \bibnamefont {Wang}},\ }\href {\doibase 10.1021/nn4038145} {\bibfield
  {journal} {\bibinfo  {journal} {ACS Nano}\ }\textbf {\bibinfo {volume} {7}},\
  \bibinfo {pages} {9205} (\bibinfo {year} {2013})}\BibitemShut {NoStop}%
\bibitem [{\citenamefont {Chang}\ \emph
  {et~al.}(2013{\natexlab{b}})\citenamefont {Chang}, \citenamefont {Zhang},
  \citenamefont {Feng}, \citenamefont {Shen}, \citenamefont {Zhang},
  \citenamefont {Guo}, \citenamefont {Li}, \citenamefont {Ou}, \citenamefont
  {Wei}, \citenamefont {Wang}, \citenamefont {Ji}, \citenamefont {Feng},
  \citenamefont {Ji}, \citenamefont {Chen}, \citenamefont {Jia}, \citenamefont
  {Dai}, \citenamefont {Fang}, \citenamefont {Zhang}, \citenamefont {He},
  \citenamefont {Wang}, \citenamefont {Lu}, \citenamefont {Ma},\ and\
  \citenamefont {Xue}}]{Tsinghua}%
  \BibitemOpen
  \bibfield  {author} {\bibinfo {author} {\bibfnamefont {C.-Z.}\ \bibnamefont
  {Chang}}, \bibinfo {author} {\bibfnamefont {J.}~\bibnamefont {Zhang}},
  \bibinfo {author} {\bibfnamefont {X.}~\bibnamefont {Feng}}, \bibinfo {author}
  {\bibfnamefont {J.}~\bibnamefont {Shen}}, \bibinfo {author} {\bibfnamefont
  {Z.}~\bibnamefont {Zhang}}, \bibinfo {author} {\bibfnamefont
  {M.}~\bibnamefont {Guo}}, \bibinfo {author} {\bibfnamefont {K.}~\bibnamefont
  {Li}}, \bibinfo {author} {\bibfnamefont {Y.}~\bibnamefont {Ou}}, \bibinfo
  {author} {\bibfnamefont {P.}~\bibnamefont {Wei}}, \bibinfo {author}
  {\bibfnamefont {L.-L.}\ \bibnamefont {Wang}}, \bibinfo {author}
  {\bibfnamefont {Z.-Q.}\ \bibnamefont {Ji}}, \bibinfo {author} {\bibfnamefont
  {Y.}~\bibnamefont {Feng}}, \bibinfo {author} {\bibfnamefont {S.}~\bibnamefont
  {Ji}}, \bibinfo {author} {\bibfnamefont {X.}~\bibnamefont {Chen}}, \bibinfo
  {author} {\bibfnamefont {J.}~\bibnamefont {Jia}}, \bibinfo {author}
  {\bibfnamefont {X.}~\bibnamefont {Dai}}, \bibinfo {author} {\bibfnamefont
  {Z.}~\bibnamefont {Fang}}, \bibinfo {author} {\bibfnamefont {S.-C.}\
  \bibnamefont {Zhang}}, \bibinfo {author} {\bibfnamefont {K.}~\bibnamefont
  {He}}, \bibinfo {author} {\bibfnamefont {Y.}~\bibnamefont {Wang}}, \bibinfo
  {author} {\bibfnamefont {L.}~\bibnamefont {Lu}}, \bibinfo {author}
  {\bibfnamefont {X.-C.}\ \bibnamefont {Ma}}, \ and\ \bibinfo {author}
  {\bibfnamefont {Q.-K.}\ \bibnamefont {Xue}},\ }\href {\doibase
  10.1126/science.1234414} {\bibfield  {journal} {\bibinfo  {journal}
  {Science}\ }\textbf {\bibinfo {volume} {340}},\ \bibinfo {pages} {167}
  (\bibinfo {year} {2013}{\natexlab{b}})}\BibitemShut {NoStop}%
\bibitem [{\citenamefont {Kou}\ \emph {et~al.}(2014)\citenamefont {Kou},
  \citenamefont {Guo}, \citenamefont {Fan}, \citenamefont {Pan}, \citenamefont
  {Lang}, \citenamefont {Jiang}, \citenamefont {Shao}, \citenamefont {Nie},
  \citenamefont {Murata}, \citenamefont {Tang}, \citenamefont {Wang},
  \citenamefont {He}, \citenamefont {Lee}, \citenamefont {Lee},\ and\
  \citenamefont {Wang}}]{UCLA}%
  \BibitemOpen
  \bibfield  {author} {\bibinfo {author} {\bibfnamefont {X.}~\bibnamefont
  {Kou}}, \bibinfo {author} {\bibfnamefont {S.-T.}\ \bibnamefont {Guo}},
  \bibinfo {author} {\bibfnamefont {Y.}~\bibnamefont {Fan}}, \bibinfo {author}
  {\bibfnamefont {L.}~\bibnamefont {Pan}}, \bibinfo {author} {\bibfnamefont
  {M.}~\bibnamefont {Lang}}, \bibinfo {author} {\bibfnamefont {Y.}~\bibnamefont
  {Jiang}}, \bibinfo {author} {\bibfnamefont {Q.}~\bibnamefont {Shao}},
  \bibinfo {author} {\bibfnamefont {T.}~\bibnamefont {Nie}}, \bibinfo {author}
  {\bibfnamefont {K.}~\bibnamefont {Murata}}, \bibinfo {author} {\bibfnamefont
  {J.}~\bibnamefont {Tang}}, \bibinfo {author} {\bibfnamefont {Y.}~\bibnamefont
  {Wang}}, \bibinfo {author} {\bibfnamefont {L.}~\bibnamefont {He}}, \bibinfo
  {author} {\bibfnamefont {T.-K.}\ \bibnamefont {Lee}}, \bibinfo {author}
  {\bibfnamefont {W.-L.}\ \bibnamefont {Lee}}, \ and\ \bibinfo {author}
  {\bibfnamefont {K.~L.}\ \bibnamefont {Wang}},\ }\href {\doibase
  10.1103/PhysRevLett.113.137201} {\bibfield  {journal} {\bibinfo  {journal}
  {Phys. Rev. Lett.}\ }\textbf {\bibinfo {volume} {113}},\ \bibinfo {pages}
  {137201} (\bibinfo {year} {2014})}\BibitemShut {NoStop}%
\bibitem [{\citenamefont {Checkelsky}\ \emph {et~al.}(2014)\citenamefont
  {Checkelsky}, \citenamefont {Yoshimi}, \citenamefont {Tsukazaki},
  \citenamefont {Takahashi}, \citenamefont {Kozuka}, \citenamefont {Falson},
  \citenamefont {Kawasaki},\ and\ \citenamefont {Tokura}}]{Tokura}%
  \BibitemOpen
  \bibfield  {author} {\bibinfo {author} {\bibfnamefont {J.~G.}\ \bibnamefont
  {Checkelsky}}, \bibinfo {author} {\bibfnamefont {R.}~\bibnamefont {Yoshimi}},
  \bibinfo {author} {\bibfnamefont {A.}~\bibnamefont {Tsukazaki}}, \bibinfo
  {author} {\bibfnamefont {K.~S.}\ \bibnamefont {Takahashi}}, \bibinfo {author}
  {\bibfnamefont {Y.}~\bibnamefont {Kozuka}}, \bibinfo {author} {\bibfnamefont
  {J.}~\bibnamefont {Falson}}, \bibinfo {author} {\bibfnamefont
  {M.}~\bibnamefont {Kawasaki}}, \ and\ \bibinfo {author} {\bibfnamefont
  {Y.}~\bibnamefont {Tokura}},\ }\href {http://dx.doi.org/10.1038/nphys3053}
  {\bibfield  {journal} {\bibinfo  {journal} {Nat. Phys.}\ }\textbf {\bibinfo
  {volume} {10}},\ \bibinfo {pages} {731} (\bibinfo {year} {2014})}\BibitemShut
  {NoStop}%
\bibitem [{\citenamefont {Wang}\ \emph
  {et~al.}(2013{\natexlab{b}})\citenamefont {Wang}, \citenamefont {Lian},
  \citenamefont {Zhang},\ and\ \citenamefont {Zhang}}]{Jing}%
  \BibitemOpen
  \bibfield  {author} {\bibinfo {author} {\bibfnamefont {J.}~\bibnamefont
  {Wang}}, \bibinfo {author} {\bibfnamefont {B.}~\bibnamefont {Lian}}, \bibinfo
  {author} {\bibfnamefont {H.}~\bibnamefont {Zhang}}, \ and\ \bibinfo {author}
  {\bibfnamefont {S.-C.}\ \bibnamefont {Zhang}},\ }\href {\doibase
  10.1103/PhysRevLett.111.086803} {\bibfield  {journal} {\bibinfo  {journal}
  {Phys. Rev. Lett.}\ }\textbf {\bibinfo {volume} {111}},\ \bibinfo {pages}
  {086803} (\bibinfo {year} {2013}{\natexlab{b}})}\BibitemShut {NoStop}%
\bibitem [{Sup()}]{SuppInfo}%
  \BibitemOpen
  \href@noop {} {}\bibinfo {note} {See supplementary information.}\BibitemShut
  {Stop}%
\bibitem [{\citenamefont {Roth}\ \emph {et~al.}(2009)\citenamefont {Roth},
  \citenamefont {Br\"une}, \citenamefont {Buhmann}, \citenamefont {Molenkamp},
  \citenamefont {Maciejko}, \citenamefont {Qi},\ and\ \citenamefont
  {Zhang}}]{MolenkampHbars}%
  \BibitemOpen
  \bibfield  {author} {\bibinfo {author} {\bibfnamefont {A.}~\bibnamefont
  {Roth}}, \bibinfo {author} {\bibfnamefont {C.}~\bibnamefont {Br\"une}},
  \bibinfo {author} {\bibfnamefont {H.}~\bibnamefont {Buhmann}}, \bibinfo
  {author} {\bibfnamefont {L.~W.}\ \bibnamefont {Molenkamp}}, \bibinfo {author}
  {\bibfnamefont {J.}~\bibnamefont {Maciejko}}, \bibinfo {author}
  {\bibfnamefont {X.-L.}\ \bibnamefont {Qi}}, \ and\ \bibinfo {author}
  {\bibfnamefont {S.-C.}\ \bibnamefont {Zhang}},\ }\href {\doibase
  10.1126/science.1174736} {\bibfield  {journal} {\bibinfo  {journal}
  {Science}\ }\textbf {\bibinfo {volume} {325}},\ \bibinfo {pages} {294}
  (\bibinfo {year} {2009})}\BibitemShut {NoStop}%
\bibitem [{\citenamefont {B\"uttiker}(1988)}]{Buttiker}%
  \BibitemOpen
  \bibfield  {author} {\bibinfo {author} {\bibfnamefont {M.}~\bibnamefont
  {B\"uttiker}},\ }\href {\doibase 10.1103/PhysRevB.38.9375} {\bibfield
  {journal} {\bibinfo  {journal} {Phys. Rev. B}\ }\textbf {\bibinfo {volume}
  {38}},\ \bibinfo {pages} {9375} (\bibinfo {year} {1988})}\BibitemShut
  {NoStop}%
\bibitem [{\citenamefont {Beenakker}\ and\ \citenamefont {van
  Houten}(1991)}]{Beenakker}%
  \BibitemOpen
  \bibfield  {author} {\bibinfo {author} {\bibfnamefont {C.}~\bibnamefont
  {Beenakker}}\ and\ \bibinfo {author} {\bibfnamefont {H.}~\bibnamefont {van
  Houten}},\ }in\ \href {\doibase
  http://dx.doi.org/10.1016/S0081-1947(08)60091-0} {\emph {\bibinfo {booktitle}
  {Semiconductor Heterostructures and Nanostructures}}},\ \bibinfo {series}
  {Solid State Physics}, Vol.~\bibinfo {volume} {44},\ \bibinfo {editor}
  {edited by\ \bibinfo {editor} {\bibfnamefont {H.}~\bibnamefont {Ehrenreich}}\
  and\ \bibinfo {editor} {\bibfnamefont {D.}~\bibnamefont {Turnbull}}}\
  (\bibinfo  {publisher} {Academic Press},\ \bibinfo {year} {1991})\ pp.\
  \bibinfo {pages} {1 -- 228}\BibitemShut {NoStop}%
\bibitem [{\citenamefont {Dolan}(1999)}]{RGflowlines}%
  \BibitemOpen
  \bibfield  {author} {\bibinfo {author} {\bibfnamefont {B.~P.}\ \bibnamefont
  {Dolan}},\ }\href {\doibase http://dx.doi.org/10.1016/S0550-3213(99)00326-0}
  {\bibfield  {journal} {\bibinfo  {journal} {Nucl. Phys. B}\ }\textbf
  {\bibinfo {volume} {554}},\ \bibinfo {pages} {487 } (\bibinfo {year}
  {1999})}\BibitemShut {NoStop}%
\bibitem [{\citenamefont {Pecharsky}\ \emph {et~al.}(2001)\citenamefont
  {Pecharsky}, \citenamefont {Gschneidner}, \citenamefont {Pecharsky},\ and\
  \citenamefont {Tishin}}]{MCE}%
  \BibitemOpen
  \bibfield  {author} {\bibinfo {author} {\bibfnamefont {V.~K.}\ \bibnamefont
  {Pecharsky}}, \bibinfo {author} {\bibfnamefont {K.~A.}\ \bibnamefont
  {Gschneidner}}, \bibinfo {author} {\bibfnamefont {A.~O.}\ \bibnamefont
  {Pecharsky}}, \ and\ \bibinfo {author} {\bibfnamefont {A.~M.}\ \bibnamefont
  {Tishin}},\ }\href {\doibase 10.1103/PhysRevB.64.144406} {\bibfield
  {journal} {\bibinfo  {journal} {Phys. Rev. B}\ }\textbf {\bibinfo {volume}
  {64}},\ \bibinfo {pages} {144406} (\bibinfo {year} {2001})}\BibitemShut
  {NoStop}%
\bibitem [{\citenamefont {{Chang}}\ \emph {et~al.}(2014)\citenamefont
  {{Chang}}, \citenamefont {{Zhao}}, \citenamefont {{Kim}}, \citenamefont
  {{Zhang}}, \citenamefont {{Assaf}}, \citenamefont {{Heiman}}, \citenamefont
  {{Zhang}}, \citenamefont {{Liu}}, \citenamefont {{Chan}},\ and\ \citenamefont
  {{Moodera}}}]{MosesChan}%
  \BibitemOpen
  \bibfield  {author} {\bibinfo {author} {\bibfnamefont {C.-Z.}\ \bibnamefont
  {{Chang}}}, \bibinfo {author} {\bibfnamefont {W.}~\bibnamefont {{Zhao}}},
  \bibinfo {author} {\bibfnamefont {D.~Y.}\ \bibnamefont {{Kim}}}, \bibinfo
  {author} {\bibfnamefont {H.}~\bibnamefont {{Zhang}}}, \bibinfo {author}
  {\bibfnamefont {B.~A.}\ \bibnamefont {{Assaf}}}, \bibinfo {author}
  {\bibfnamefont {D.}~\bibnamefont {{Heiman}}}, \bibinfo {author}
  {\bibfnamefont {S.-C.}\ \bibnamefont {{Zhang}}}, \bibinfo {author}
  {\bibfnamefont {C.}~\bibnamefont {{Liu}}}, \bibinfo {author} {\bibfnamefont
  {M.~H.~W.}\ \bibnamefont {{Chan}}}, \ and\ \bibinfo {author} {\bibfnamefont
  {J.~S.}\ \bibnamefont {{Moodera}}},\ }\href@noop {} {\bibfield  {journal}
  {\bibinfo  {journal} {ArXiv e-prints}\ } (\bibinfo {year} {2014})},\ \Eprint
  {http://arxiv.org/abs/1412.3758} {arXiv:1412.3758 [cond-mat.mtrl-sci]}
  \BibitemShut {NoStop}%
\end{thebibliography}%


\begin{thebibliography}{12}%
\makeatletter
\providecommand \@ifxundefined [1]{%
 \@ifx{#1\undefined}
}%
\providecommand \@ifnum [1]{%
 \ifnum #1\expandafter \@firstoftwo
 \else \expandafter \@secondoftwo
 \fi
}%
\providecommand \@ifx [1]{%
 \ifx #1\expandafter \@firstoftwo
 \else \expandafter \@secondoftwo
 \fi
}%
\providecommand \natexlab [1]{#1}%
\providecommand \enquote  [1]{``#1''}%
\providecommand \bibnamefont  [1]{#1}%
\providecommand \bibfnamefont [1]{#1}%
\providecommand \citenamefont [1]{#1}%
\providecommand \href@noop [0]{\@secondoftwo}%
\providecommand \href [0]{\begingroup \@sanitize@url \@href}%
\providecommand \@href[1]{\@@startlink{#1}\@@href}%
\providecommand \@@href[1]{\endgroup#1\@@endlink}%
\providecommand \@sanitize@url [0]{\catcode `\\12\catcode `\$12\catcode
  `\&12\catcode `\#12\catcode `\^12\catcode `\_12\catcode `\%12\relax}%
\providecommand \@@startlink[1]{}%
\providecommand \@@endlink[0]{}%
\providecommand \url  [0]{\begingroup\@sanitize@url \@url }%
\providecommand \@url [1]{\endgroup\@href {#1}{\urlprefix }}%
\providecommand \urlprefix  [0]{URL }%
\providecommand \Eprint [0]{\href }%
\providecommand \doibase [0]{http://dx.doi.org/}%
\providecommand \selectlanguage [0]{\@gobble}%
\providecommand \bibinfo  [0]{\@secondoftwo}%
\providecommand \bibfield  [0]{\@secondoftwo}%
\providecommand \translation [1]{[#1]}%
\providecommand \BibitemOpen [0]{}%
\providecommand \bibitemStop [0]{}%
\providecommand \bibitemNoStop [0]{.\EOS\space}%
\providecommand \EOS [0]{\spacefactor3000\relax}%
\providecommand \BibitemShut  [1]{\csname bibitem#1\endcsname}%
\let\auto@bib@innerbib\@empty
\bibitem [{\citenamefont {He}\ \emph {et~al.}(2013)\citenamefont {He},
  \citenamefont {Kou}, \citenamefont {Lang}, \citenamefont {Choi},
  \citenamefont {Jiang}, \citenamefont {Nie}, \citenamefont {Jiang},
  \citenamefont {Fan}, \citenamefont {Wang}, \citenamefont {Xiu},\ and\
  \citenamefont {Wang}}]{SUCLA2}%
  \BibitemOpen
  \bibfield  {author} {\bibinfo {author} {\bibfnamefont {L.}~\bibnamefont
  {He}}, \bibinfo {author} {\bibfnamefont {X.}~\bibnamefont {Kou}}, \bibinfo
  {author} {\bibfnamefont {M.}~\bibnamefont {Lang}}, \bibinfo {author}
  {\bibfnamefont {E.~S.}\ \bibnamefont {Choi}}, \bibinfo {author}
  {\bibfnamefont {Y.}~\bibnamefont {Jiang}}, \bibinfo {author} {\bibfnamefont
  {T.}~\bibnamefont {Nie}}, \bibinfo {author} {\bibfnamefont {W.}~\bibnamefont
  {Jiang}}, \bibinfo {author} {\bibfnamefont {Y.}~\bibnamefont {Fan}}, \bibinfo
  {author} {\bibfnamefont {Y.}~\bibnamefont {Wang}}, \bibinfo {author}
  {\bibfnamefont {F.}~\bibnamefont {Xiu}}, \ and\ \bibinfo {author}
  {\bibfnamefont {K.~L.}\ \bibnamefont {Wang}},\ }\href {\doibase
  http://dx.doi.org/10.1038/srep03406} {\bibfield  {journal} {\bibinfo
  {journal} {Sci. Rep.}\ }\textbf {\bibinfo {volume} {3}},\ \bibinfo {pages}
  {3406} (\bibinfo {year} {2013})}\BibitemShut {NoStop}%
\bibitem [{\citenamefont {Lang}\ \emph {et~al.}(2012)\citenamefont {Lang},
  \citenamefont {He}, \citenamefont {Xiu}, \citenamefont {Yu}, \citenamefont
  {Tang}, \citenamefont {Wang}, \citenamefont {Kou}, \citenamefont {Jiang},
  \citenamefont {Fedorov},\ and\ \citenamefont {Wang}}]{SUCLA3}%
  \BibitemOpen
  \bibfield  {author} {\bibinfo {author} {\bibfnamefont {M.}~\bibnamefont
  {Lang}}, \bibinfo {author} {\bibfnamefont {L.}~\bibnamefont {He}}, \bibinfo
  {author} {\bibfnamefont {F.}~\bibnamefont {Xiu}}, \bibinfo {author}
  {\bibfnamefont {X.}~\bibnamefont {Yu}}, \bibinfo {author} {\bibfnamefont
  {J.}~\bibnamefont {Tang}}, \bibinfo {author} {\bibfnamefont {Y.}~\bibnamefont
  {Wang}}, \bibinfo {author} {\bibfnamefont {X.}~\bibnamefont {Kou}}, \bibinfo
  {author} {\bibfnamefont {W.}~\bibnamefont {Jiang}}, \bibinfo {author}
  {\bibfnamefont {A.~V.}\ \bibnamefont {Fedorov}}, \ and\ \bibinfo {author}
  {\bibfnamefont {K.~L.}\ \bibnamefont {Wang}},\ }\href {\doibase
  10.1021/nn204239d} {\bibfield  {journal} {\bibinfo  {journal} {ACS Nano}\
  }\textbf {\bibinfo {volume} {6}},\ \bibinfo {pages} {295} (\bibinfo {year}
  {2012})}\BibitemShut {NoStop}%
\bibitem [{\citenamefont {Milliken}\ \emph {et~al.}(2007)\citenamefont
  {Milliken}, \citenamefont {Rozen}, \citenamefont {Keefe},\ and\ \citenamefont
  {Koch}}]{Sfilters}%
  \BibitemOpen
  \bibfield  {author} {\bibinfo {author} {\bibfnamefont {F.~P.}\ \bibnamefont
  {Milliken}}, \bibinfo {author} {\bibfnamefont {J.~R.}\ \bibnamefont {Rozen}},
  \bibinfo {author} {\bibfnamefont {G.~A.}\ \bibnamefont {Keefe}}, \ and\
  \bibinfo {author} {\bibfnamefont {R.~H.}\ \bibnamefont {Koch}},\ }\href
  {\doibase http://dx.doi.org/10.1063/1.2431770} {\bibfield  {journal}
  {\bibinfo  {journal} {Rev. Sci. Instrum.}\ }\textbf {\bibinfo {volume}
  {78}},\ \bibinfo {eid} {024701} (\bibinfo {year} {2007})}\BibitemShut
  {NoStop}%
\bibitem [{\citenamefont {Fischer}\ and\ \citenamefont
  {Grayson}(2005)}]{Sinputimpedance}%
  \BibitemOpen
  \bibfield  {author} {\bibinfo {author} {\bibfnamefont {F.}~\bibnamefont
  {Fischer}}\ and\ \bibinfo {author} {\bibfnamefont {M.}~\bibnamefont
  {Grayson}},\ }\href {\doibase http://dx.doi.org/10.1063/1.1948530} {\bibfield
   {journal} {\bibinfo  {journal} {J. Appl. Phys.}\ }\textbf {\bibinfo {volume}
  {98}},\ \bibinfo {eid} {013710} (\bibinfo {year} {2005})}\BibitemShut
  {NoStop}%
\bibitem [{\citenamefont {B\"uttiker}(1988)}]{SButtiker}%
  \BibitemOpen
  \bibfield  {author} {\bibinfo {author} {\bibfnamefont {M.}~\bibnamefont
  {B\"uttiker}},\ }\href {\doibase 10.1103/PhysRevB.38.9375} {\bibfield
  {journal} {\bibinfo  {journal} {Phys. Rev. B}\ }\textbf {\bibinfo {volume}
  {38}},\ \bibinfo {pages} {9375} (\bibinfo {year} {1988})}\BibitemShut
  {NoStop}%
\bibitem [{\citenamefont {Kou}\ \emph {et~al.}(2014)\citenamefont {Kou},
  \citenamefont {Guo}, \citenamefont {Fan}, \citenamefont {Pan}, \citenamefont
  {Lang}, \citenamefont {Jiang}, \citenamefont {Shao}, \citenamefont {Nie},
  \citenamefont {Murata}, \citenamefont {Tang}, \citenamefont {Wang},
  \citenamefont {He}, \citenamefont {Lee}, \citenamefont {Lee},\ and\
  \citenamefont {Wang}}]{SUCLA}%
  \BibitemOpen
  \bibfield  {author} {\bibinfo {author} {\bibfnamefont {X.}~\bibnamefont
  {Kou}}, \bibinfo {author} {\bibfnamefont {S.-T.}\ \bibnamefont {Guo}},
  \bibinfo {author} {\bibfnamefont {Y.}~\bibnamefont {Fan}}, \bibinfo {author}
  {\bibfnamefont {L.}~\bibnamefont {Pan}}, \bibinfo {author} {\bibfnamefont
  {M.}~\bibnamefont {Lang}}, \bibinfo {author} {\bibfnamefont {Y.}~\bibnamefont
  {Jiang}}, \bibinfo {author} {\bibfnamefont {Q.}~\bibnamefont {Shao}},
  \bibinfo {author} {\bibfnamefont {T.}~\bibnamefont {Nie}}, \bibinfo {author}
  {\bibfnamefont {K.}~\bibnamefont {Murata}}, \bibinfo {author} {\bibfnamefont
  {J.}~\bibnamefont {Tang}}, \bibinfo {author} {\bibfnamefont {Y.}~\bibnamefont
  {Wang}}, \bibinfo {author} {\bibfnamefont {L.}~\bibnamefont {He}}, \bibinfo
  {author} {\bibfnamefont {T.-K.}\ \bibnamefont {Lee}}, \bibinfo {author}
  {\bibfnamefont {W.-L.}\ \bibnamefont {Lee}}, \ and\ \bibinfo {author}
  {\bibfnamefont {K.~L.}\ \bibnamefont {Wang}},\ }\href {\doibase
  10.1103/PhysRevLett.113.137201} {\bibfield  {journal} {\bibinfo  {journal}
  {Phys. Rev. Lett.}\ }\textbf {\bibinfo {volume} {113}},\ \bibinfo {pages}
  {137201} (\bibinfo {year} {2014})}\BibitemShut {NoStop}%
\bibitem [{\citenamefont {Tsui}\ \emph {et~al.}(1982)\citenamefont {Tsui},
  \citenamefont {Stormer},\ and\ \citenamefont {Gossard}}]{SArrhenius1}%
  \BibitemOpen
  \bibfield  {author} {\bibinfo {author} {\bibfnamefont {D.~C.}\ \bibnamefont
  {Tsui}}, \bibinfo {author} {\bibfnamefont {H.~L.}\ \bibnamefont {Stormer}}, \
  and\ \bibinfo {author} {\bibfnamefont {A.~C.}\ \bibnamefont {Gossard}},\
  }\href {\doibase 10.1103/PhysRevLett.48.1559} {\bibfield  {journal} {\bibinfo
   {journal} {Phys. Rev. Lett.}\ }\textbf {\bibinfo {volume} {48}},\ \bibinfo
  {pages} {1559} (\bibinfo {year} {1982})}\BibitemShut {NoStop}%
\bibitem [{\citenamefont {Gammel}\ \emph {et~al.}(1988)\citenamefont {Gammel},
  \citenamefont {Bishop}, \citenamefont {Eisenstein}, \citenamefont {English},
  \citenamefont {Gossard}, \citenamefont {Ruel},\ and\ \citenamefont
  {Stormer}}]{SArrhenius2}%
  \BibitemOpen
  \bibfield  {author} {\bibinfo {author} {\bibfnamefont {P.~L.}\ \bibnamefont
  {Gammel}}, \bibinfo {author} {\bibfnamefont {D.~J.}\ \bibnamefont {Bishop}},
  \bibinfo {author} {\bibfnamefont {J.~P.}\ \bibnamefont {Eisenstein}},
  \bibinfo {author} {\bibfnamefont {J.~H.}\ \bibnamefont {English}}, \bibinfo
  {author} {\bibfnamefont {A.~C.}\ \bibnamefont {Gossard}}, \bibinfo {author}
  {\bibfnamefont {R.}~\bibnamefont {Ruel}}, \ and\ \bibinfo {author}
  {\bibfnamefont {H.~L.}\ \bibnamefont {Stormer}},\ }\href {\doibase
  10.1103/PhysRevB.38.10128} {\bibfield  {journal} {\bibinfo  {journal} {Phys.
  Rev. B}\ }\textbf {\bibinfo {volume} {38}},\ \bibinfo {pages} {10128}
  (\bibinfo {year} {1988})}\BibitemShut {NoStop}%
\bibitem [{\citenamefont {Chang}\ \emph {et~al.}(2013)\citenamefont {Chang},
  \citenamefont {Zhang}, \citenamefont {Feng}, \citenamefont {Shen},
  \citenamefont {Zhang}, \citenamefont {Guo}, \citenamefont {Li}, \citenamefont
  {Ou}, \citenamefont {Wei}, \citenamefont {Wang}, \citenamefont {Ji},
  \citenamefont {Feng}, \citenamefont {Ji}, \citenamefont {Chen}, \citenamefont
  {Jia}, \citenamefont {Dai}, \citenamefont {Fang}, \citenamefont {Zhang},
  \citenamefont {He}, \citenamefont {Wang}, \citenamefont {Lu}, \citenamefont
  {Ma},\ and\ \citenamefont {Xue}}]{STsinghua}%
  \BibitemOpen
  \bibfield  {author} {\bibinfo {author} {\bibfnamefont {C.-Z.}\ \bibnamefont
  {Chang}}, \bibinfo {author} {\bibfnamefont {J.}~\bibnamefont {Zhang}},
  \bibinfo {author} {\bibfnamefont {X.}~\bibnamefont {Feng}}, \bibinfo {author}
  {\bibfnamefont {J.}~\bibnamefont {Shen}}, \bibinfo {author} {\bibfnamefont
  {Z.}~\bibnamefont {Zhang}}, \bibinfo {author} {\bibfnamefont
  {M.}~\bibnamefont {Guo}}, \bibinfo {author} {\bibfnamefont {K.}~\bibnamefont
  {Li}}, \bibinfo {author} {\bibfnamefont {Y.}~\bibnamefont {Ou}}, \bibinfo
  {author} {\bibfnamefont {P.}~\bibnamefont {Wei}}, \bibinfo {author}
  {\bibfnamefont {L.-L.}\ \bibnamefont {Wang}}, \bibinfo {author}
  {\bibfnamefont {Z.-Q.}\ \bibnamefont {Ji}}, \bibinfo {author} {\bibfnamefont
  {Y.}~\bibnamefont {Feng}}, \bibinfo {author} {\bibfnamefont {S.}~\bibnamefont
  {Ji}}, \bibinfo {author} {\bibfnamefont {X.}~\bibnamefont {Chen}}, \bibinfo
  {author} {\bibfnamefont {J.}~\bibnamefont {Jia}}, \bibinfo {author}
  {\bibfnamefont {X.}~\bibnamefont {Dai}}, \bibinfo {author} {\bibfnamefont
  {Z.}~\bibnamefont {Fang}}, \bibinfo {author} {\bibfnamefont {S.-C.}\
  \bibnamefont {Zhang}}, \bibinfo {author} {\bibfnamefont {K.}~\bibnamefont
  {He}}, \bibinfo {author} {\bibfnamefont {Y.}~\bibnamefont {Wang}}, \bibinfo
  {author} {\bibfnamefont {L.}~\bibnamefont {Lu}}, \bibinfo {author}
  {\bibfnamefont {X.-C.}\ \bibnamefont {Ma}}, \ and\ \bibinfo {author}
  {\bibfnamefont {Q.-K.}\ \bibnamefont {Xue}},\ }\href {\doibase
  10.1126/science.1234414} {\bibfield  {journal} {\bibinfo  {journal}
  {Science}\ }\textbf {\bibinfo {volume} {340}},\ \bibinfo {pages} {167}
  (\bibinfo {year} {2013})}\BibitemShut {NoStop}%
\bibitem [{\citenamefont {Chang}\ \emph {et~al.}(2014)\citenamefont {Chang},
  \citenamefont {Tang}, \citenamefont {Wang}, \citenamefont {Feng},
  \citenamefont {Li}, \citenamefont {Zhang}, \citenamefont {Wang},
  \citenamefont {Wang}, \citenamefont {Chen}, \citenamefont {Liu},
  \citenamefont {Duan}, \citenamefont {He}, \citenamefont {Ma},\ and\
  \citenamefont {Xue}}]{Sclusters}%
  \BibitemOpen
  \bibfield  {author} {\bibinfo {author} {\bibfnamefont {C.-Z.}\ \bibnamefont
  {Chang}}, \bibinfo {author} {\bibfnamefont {P.}~\bibnamefont {Tang}},
  \bibinfo {author} {\bibfnamefont {Y.-L.}\ \bibnamefont {Wang}}, \bibinfo
  {author} {\bibfnamefont {X.}~\bibnamefont {Feng}}, \bibinfo {author}
  {\bibfnamefont {K.}~\bibnamefont {Li}}, \bibinfo {author} {\bibfnamefont
  {Z.}~\bibnamefont {Zhang}}, \bibinfo {author} {\bibfnamefont
  {Y.}~\bibnamefont {Wang}}, \bibinfo {author} {\bibfnamefont {L.-L.}\
  \bibnamefont {Wang}}, \bibinfo {author} {\bibfnamefont {X.}~\bibnamefont
  {Chen}}, \bibinfo {author} {\bibfnamefont {C.}~\bibnamefont {Liu}}, \bibinfo
  {author} {\bibfnamefont {W.}~\bibnamefont {Duan}}, \bibinfo {author}
  {\bibfnamefont {K.}~\bibnamefont {He}}, \bibinfo {author} {\bibfnamefont
  {X.-C.}\ \bibnamefont {Ma}}, \ and\ \bibinfo {author} {\bibfnamefont {Q.-K.}\
  \bibnamefont {Xue}},\ }\href {\doibase 10.1103/PhysRevLett.112.056801}
  {\bibfield  {journal} {\bibinfo  {journal} {Phys. Rev. Lett.}\ }\textbf
  {\bibinfo {volume} {112}},\ \bibinfo {pages} {056801} (\bibinfo {year}
  {2014})}\BibitemShut {NoStop}%
\bibitem [{\citenamefont {Elliott}\ \emph {et~al.}(1984)\citenamefont
  {Elliott}, \citenamefont {Chen}, \citenamefont {Greenbaum},\ and\
  \citenamefont {Wagner}}]{SGaAspara}%
  \BibitemOpen
  \bibfield  {author} {\bibinfo {author} {\bibfnamefont {K.}~\bibnamefont
  {Elliott}}, \bibinfo {author} {\bibfnamefont {R.~T.}\ \bibnamefont {Chen}},
  \bibinfo {author} {\bibfnamefont {S.~G.}\ \bibnamefont {Greenbaum}}, \ and\
  \bibinfo {author} {\bibfnamefont {R.~J.}\ \bibnamefont {Wagner}},\ }\href
  {\doibase http://dx.doi.org/10.1063/1.94930} {\bibfield  {journal} {\bibinfo
  {journal} {Appl. Phys. Lett.}\ }\textbf {\bibinfo {volume} {44}},\ \bibinfo
  {pages} {907} (\bibinfo {year} {1984})}\BibitemShut {NoStop}%
\bibitem [{\citenamefont {Blakemore}(1982)}]{Sdebye}%
  \BibitemOpen
  \bibfield  {author} {\bibinfo {author} {\bibfnamefont {J.~S.}\ \bibnamefont
  {Blakemore}},\ }\href {\doibase http://dx.doi.org/10.1063/1.331665}
  {\bibfield  {journal} {\bibinfo  {journal} {J. Appl. Phys.}\ }\textbf
  {\bibinfo {volume} {53}},\ \bibinfo {pages} {R123} (\bibinfo {year}
  {1982})}\BibitemShut {NoStop}%
\end{thebibliography}%

\section{Supplementary information}

\renewcommand{\theequation}{S.\arabic{equation}}

\stepcounter{mysection}
\makeatletter 
\renewcommand{\thefigure}{S\@arabic\c@figure}
\makeatother

\begin{figure*}
\includegraphics[width=6in]{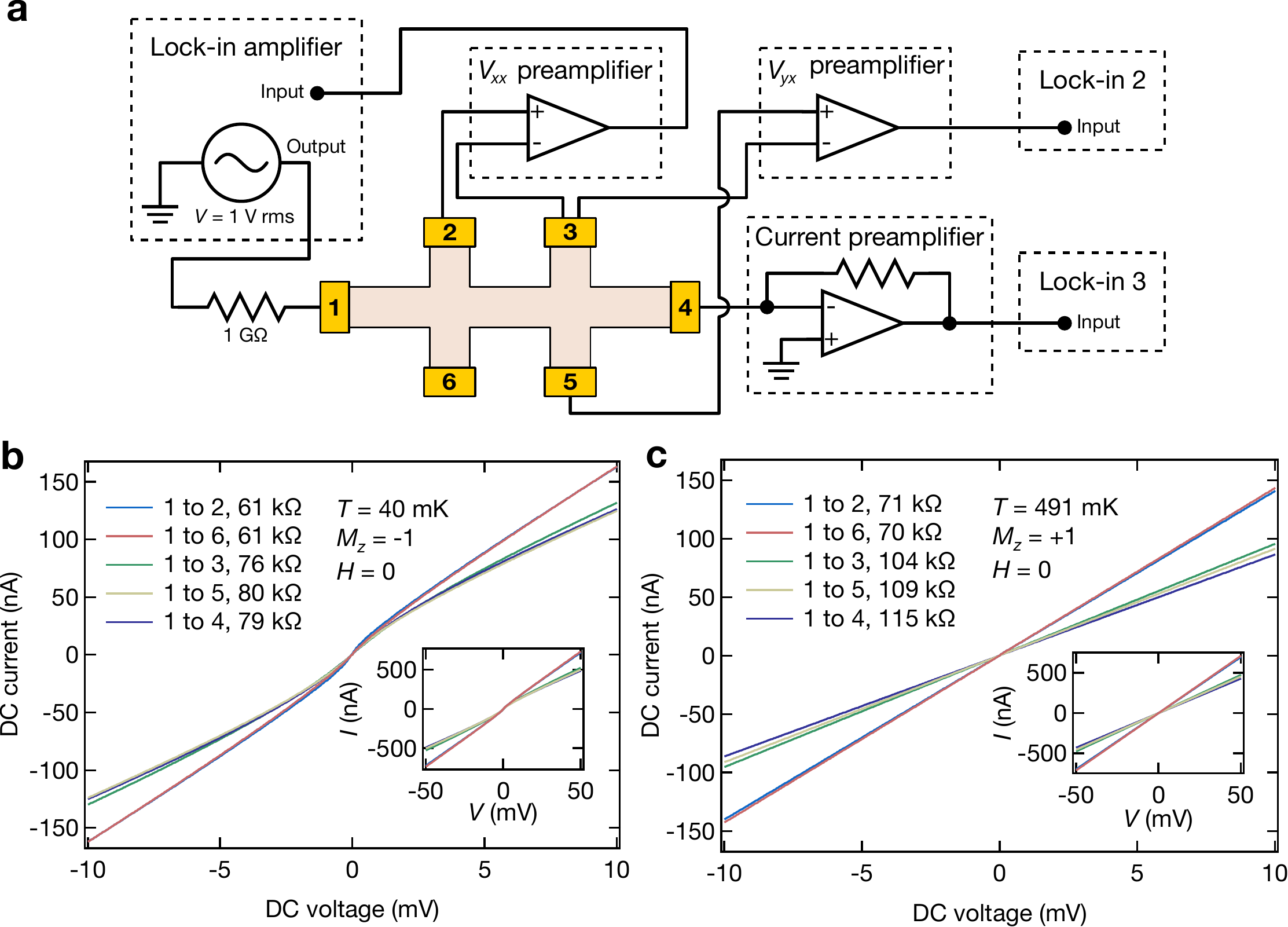}
\caption{\label{figs1} Measurement specifics. (a), Detailed schematic of Hall bar measurement. (b), (c), Two-terminal current-voltage curves at 40 mK (b) and 491 mK (c(a)). In the latter case, the indium contacts appear to be ohmic and linear, and we believe the curvature in the former case is the result of Joule heating. The insets show the same measurements over a larger voltage range.}
\end{figure*}

\begin{figure*}
\includegraphics[width=6.5in]{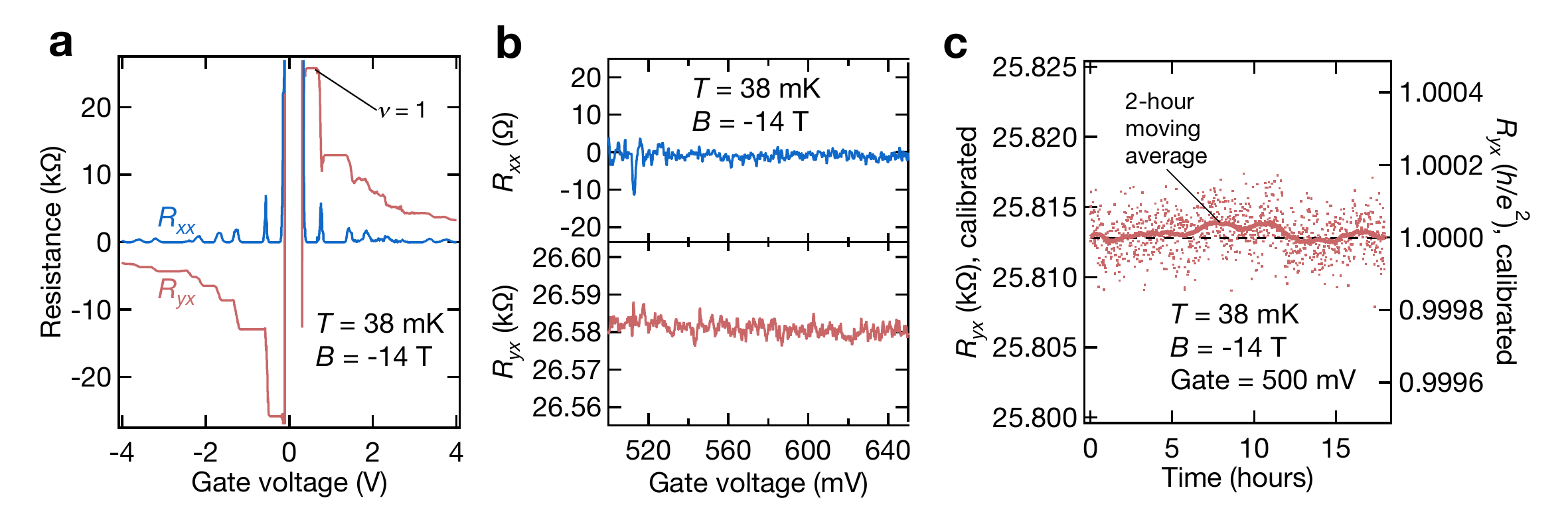}
\caption{\label{figs2} Calibration to quantum Hall plateau. (a), Integer quantum Hall effect in monolayer graphene Hall bar. (b), Uncalibrated measurement of $\nu=1$ plateau. $R_{yx}$ is found to be 2.977\% above $h/e^2$. (c), Test of calibration drift over 18-hour timescale.}
\end{figure*}

\subsection{Material growth}

High-quality single crystalline Cr-doped (Bi$_\text{x}$Sb$_\text{1-x}$)$_\text{2}$Te$_\text{3}$ films were grown in an ultra-high-vacuum Perkin-Elmer molecular beam epitaxy system. A semi-insulating ($\rho>$ 10$^\text{6}$ $\Omega\cdot$cm) GaAs (111)B substrate was cleaned by acetone in an ultrasonic bath for 10 minutes before being loaded into the growth chamber. The substrate was then annealed to 580 $^\circ$C under a Se-rich environment to remove the native oxide (as shown in Fig.\ S1(a) in~\citeS{SUCLA2}). During the growth, the GaAs substrate was maintained around 200 $^\circ$C (growth temperature), with the Bi, Sb, Te, and Cr shutters opened at the same time. Epitaxial growth was monitored by \emph{in situ} reflection high-energy electron diffraction (RHEED), and the as-grown surface configuration was traced by using the \emph{d}-spacing evolution between the two first-order diffraction lines. We observed the lattice relaxation occurring within the first QL, indicating that the surface transition from the pristine GaAs to the Cr-doped (Bi$_\text{x}$Sb$_\text{1-x}$)$_\text{2}$Te$_\text{3}$ completed immediately after the formation of the first quintuple layer. After the film growth, 2 nm Al was \emph{in situ} evaporated (for 4 minutes at 5 \AA/min) to passivate the surface at room temperature. The Al film was later naturally oxidized to form Al$_\text{2}$O$_\text{3}$ after the sample was taken out of the chamber, with the resulting oxide layer effectively protecting the grown magnetic TI film from unwanted environmental doping or other possible aging effects~\citeS{SUCLA3}. Previous atomic force microscope characterizations of the film surface demonstrated flat regions with quintuple layer steps both with and without this cap layer, suggesting the alumina coverage is complete and uniform~\citeS{SUCLA3}.

\subsection{Measurement setup}

A more detailed schematic of the measurement is shown in Figure \ref{figs1}(a). We apply a 1 nA AC bias to the device by sourcing 1 V RMS across a 1 G$\Omega$ reference resistor. The current flows across the device and out the drain terminal to an Ithaco 1211 current preamplifier, the output signal of which reflects the magnitude of the current. The preamplifier gain is set to 1 V/10$^{-7}$ A, with a 200 $\Omega$ input impedance. The differential voltages $V_{yx}$ and $V_{xx}$ are taken with NF Corporation model LI-75A voltage preamplifiers, which have 100x gain and 100 M$\Omega$ input impedances. All three preamplifier outputs are measured with SR830 lock-in amplifiers, utilizing the offset and 10x expansion feature to avoid digital discretization in $V_{yx}$ and current. The excitation/measurement frequency is 3.889 Hz. Using higher frequencies results in large phase differences between the excitation and measurement (for instance, up to 20$^{\circ}$ at 13 Hz).

The indium contacts were tested for linearity at low temperature by taking DC current-voltage (I-V) curves with a Keithley Model 2400 Source-Measure Unit. Although the resulting I-V curves are not perfectly linear at base temperature (Fig.\ \ref{figs1}(b)), we believe this results from Joule heating and the device's sensitivity to changes in temperature. At low bias, the slopes correspond to two-terminal resistances around 30 k$\Omega$, near $h/e^2$, whereas over the full voltage range the resistances rise to 60-80 k$\Omega$. These are reasonable values if, at the high end of the bias range, the temperature rises to $\sim$200 mK, at which point $\rho_{xx}\sim0.3h/e^2$. The same measurements taken at 491 mK (Fig.\ \ref{figs1}(c)) are quite linear.

Good cryogenic electronic filtering is essential to obtain low electron temperatures. At the mixing chamber stage of our cryostat, measurement lines are filtered in the GHz range by running through a cured mixture of bronze powder and epoxy~\citeS{Sfilters}, and in the MHz range by five-pole RC filters mounted on sapphire plates to improve thermal anchoring.

\subsection{Amplifier calibration}

The amplifier chain in the measurement setup requires calibration for precision measurement: the SR830 gain is rated to $\pm$1\%, the Ithaco 1211 to $\pm$2\%, and the NF LI-75A to $\pm$1\%. As a resistance standard, we use a $\nu=1$ quantum Hall plateau (where $\nu$ is the QHE filling factor) on a separate high-mobility monolayer graphene Hall bar, cooled in the cryostat at the same time. This has the advantage of being low-noise and quantized to the same value as $\rho_{yx}$ in the QAHE system. In a -14 T magnetic field while measured in the exact same configuration as the QAHE device, a gate sweep of the graphene device yields the usual progression of integer plateaus in $R_{yx}$ (Fig.\ \ref{figs2}(a)) A higher resolution sweep indicates that the uncalibrated measurement of $R_{yx}$ in the $\nu=1$ plateau is 26.581 k$\Omega$, which is 2.977\% higher than $h/e^2$ (Fig.\ \ref{figs2}(b)). In all other data reported in this paper we correct for this inaccuracy by multiplying resistances by a factor of 0.97109.

The stability of the calibration was tested by repeatedly measuring $R_{yx}$ in the $\nu=1$ plateau over an 18-hour period. During this time the measurement drift largely remained less than 0.01\% (Fig.\ \ref{figs2}(c)).

\subsection{Current loss to voltmeters}

The finite (100 M$\Omega$) input impedance of each connection to the voltage preamplifiers provides additional paths to ground for the applied current, and can distort four-terminal measurements if not accounted for. We discuss this effect in two limits: away from quantization, when bulk/surface conduction dominates, and near quantization, when chiral edge modes dominate.

For contact $i$, let the voltage be $V_i$ and the resistance to ground be $R_i$. Based on the arrangement of preamplifiers (Fig.\ \ref{figs1}(a)), $R_2=\text{100}$ M$\Omega$, $R_3=\text{50}$ M$\Omega$, and $R_5=\text{100}$ M$\Omega$. The combination of contact resistance, line resistance, and current preamplifier input impedance sets $R_4\approx\text{1000}$ $\Omega$. The current preamplifier's reading will differ most from the actual current flowing across the Hall bar when the voltages at the various leads, and therefore the currents lost through them, are maximized. Since we apply a current bias, this happens when the film is most resistive. To make an upper approximation of the inaccuracy away from quantization, we take the largest measured two-terminal resistance, 120 k$\Omega$, in which case the 1 nA bias sets $V_2\approx\text{80}$ $\mu$V and $V_3\approx V_5\approx\text{40}$ $\mu$V. The Hall effect can also raise the voltage on one side of the device relative to the other by as much as 20 $\mu$V, leading (in the worse of the two magnetization directions) to $V_2\approx\text{90}$ $\mu$V, $V_3\approx\text{50}$ $\mu$V, $V_5\approx\text{30}$ $\mu$V. This results in a total lost current of of 2.2 pA. Hence, away from quantization these extra paths to ground lead to inaccuracies in $R_{xx}$ and $R_{xy}$ as large as 0.22\%, much worse than the part-per-10,000 accuracy we achieve at quantization.

Since we calibrate our amplifiers to a system with, in principle, identical quantization, the lost current should be taken into account at the quantization limit. However, these impedances can also induce a small voltage drop along edge channels. This has been observed in the QHE~\citeS{Sinputimpedance}: at $\nu=\pm$1, $R_{xx}$ gains an extra component $(h/e^2)^2/R_i$ at one chirality but not the other (where $R_i$ is the voltmeter input impedance). The asymmetry arises because $V_{xx}$ is measured on only side of the device. At one chirality that side's voltage will be the same as the nearly-grounded drain lead, so that negligible current flows across $R_i$. At the other chirality, the measured side is in equilibrium with the source lead and hence has a larger potential difference with respect to ground. The resulting current flowing out the voltmeter modifies the potential at each voltage lead, as derived below.

For $M_z=+$1 (counterclockwise chirality), according to Landauer-B\"uttiker formalism~\citeS{SButtiker} the current flowing into the device from contact $i$ is given by
\begin{equation*}
I_i = \frac{e^2}{h}(V_i-V_{i+1})
\end{equation*}
where we identify $i=\text{7}$ with $i=\text{1}$. We also have $V_i = -R_iI_i$ for $i=\text{2,3,4,5,}$ and $I_6=\text{0}$. Solving this system of equations yields expected measurements
\begin{widetext}
\begin{align*}
R_{xx} &= \frac{V_3-V_2}{-I_4} = -\frac{h}{e^2}\frac{R_4 R_3}{(R_2+h/e^2)(R_3+h/e^2)}\approx-\text{0.3 }\Omega,\\
R_{yx} &= \frac{V_5-V_3}{-I_4} = \frac{h}{e^2}\left(1+\frac{R_4}{R_3+h/e^2}\right)\approx \frac{h}{e^2}+\text{0.5 }\Omega.
\end{align*}
\end{widetext}
For $M_z=-$1 (clockwise chirality), we have instead
\begin{equation*}
I_i = \frac{e^2}{h}(V_i-V_{i-1})
\end{equation*}
which yields
\begin{align*}
R_{xx} &= \frac{h}{e^2}\frac{R_4+h/e^2}{R_3}\approx\text{14 }\Omega,\\
R_{yx} &=-\frac{h}{e^2}\left(1+\frac{R_4}{R_5+h/e^2}\right)\approx -\left(\frac{h}{e^2}+\text{0.3 }\Omega\right).
\end{align*}

For both the calibration and QAHE quantization measurement, the systematic error in $R_{yx}$ is within the measurement noise. We only have a sizable addition to $R_{xx}$ when $M_z=-$1, which may be one reason we obtain lower values of $\rho_{xx}$ and $\sigma_{xx}$ for $M_z=+$1 (see top and middle panels of Fig.\ \ref{fig2}(a)). The effect also has important implications for asymmetries in four-terminal non-local measurements, as discussed in a later section.

\subsection{Device dimensions}

\begin{figure*}
\includegraphics[width=6.5in]{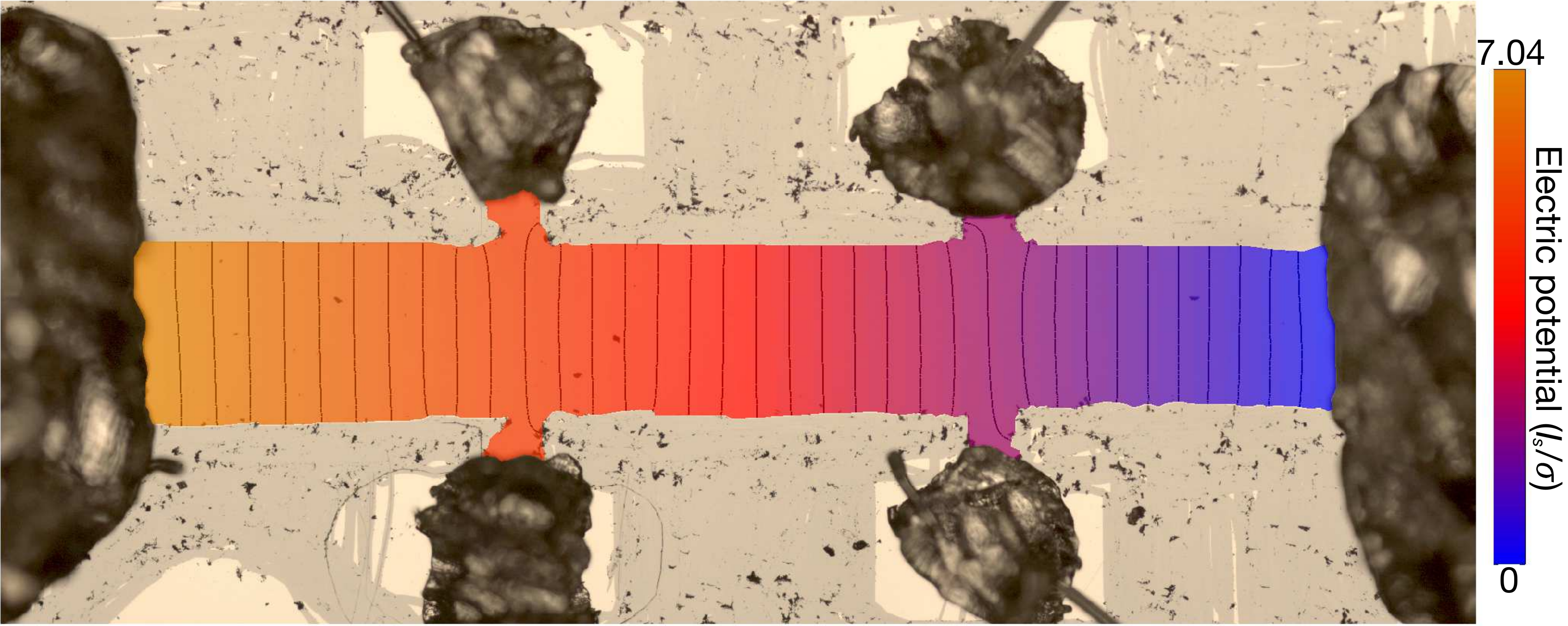}
\caption{\label{figs3} Calculated electric potential, $\varphi(\mathbf{r})$, on the Hall bar. Numerical solution for the electric potential on a polygon model of the Hall bar, in units of $I_s/\sigma$, where $I_s$ is the source current and $\sigma$ is the sheet conductivity, shown overlaid on a micrograph of the device, with a number of equipotentials (black lines). The calculation assumes a uniform sheet conductivity, with zero current flowing into the voltage leads, and finds the number of squares between the top longitudinal voltage leads to be $N_\square = $ 2.77.}
\end{figure*}

To find the diagonal resistivity, $\rho_{xx} = R_{xx}/N_\square$, from the measured four-terminal longitudinal resistance, $R_{xx}$, the number of squares, $N_\square$, between the longitudinal voltage contacts of the Hall bar must be known. However, since this device is not fabricated with precise lithography, $N_\square$ cannot be inferred from the device design. Instead, to calculate $N_\square$, we numerically solve the Laplace equation on a 2D polygon model of the device.

In a resistive material with a static current distribution, no charge accumulation, and in the absence of a magnetic field, the electric potential, $\varphi(\mathbf{r})$, satisfies Laplace equation $\nabla^2 \varphi = 0$. At the interface of the material and the vacuum, the potential satisfies the Neumann boundary condition $\frac{\partial \varphi}{\partial \mathbf{n}}\Big|_\Gamma = 0$, where $\bf{n}$ is the outward surface normal unit vector and $\Gamma$ the interface, in order that the current density perpendicular to the boundary, $\mathbf{J}\cdot \mathbf{n} = \sigma \mathbf{E}\cdot \mathbf{n} = -\sigma \frac{\partial \varphi}{\partial \mathbf{n}}$, be zero. At an interface with a perfect conductor, which must be an equipotential, the potential instead satisfies the Dirichlet boundary condition $\varphi\big|_\Gamma = V_c$, where $V_c$ is the potential of the conductor.

In the case of a Hall bar, the potential at the conductors acting as voltage leads is not fixed externally, but rather set by the condition that the net current into each voltage lead be zero. Let $\Gamma$ be the boundary between the Hall bar and the voltage lead (a curve, in the 2D case). We assume a uniform conductivity $\sigma$ for the Hall bar. Then the net current flowing into the Hall bar from the voltage lead is
\begin{equation}\label{eq:current}
I = -\int_\Gamma ds \, \mathbf{J}\cdot \mathbf{n} = \sigma \int_\Gamma ds \, \frac{\partial \varphi}{\partial \mathbf{n}},
\end{equation}
so the $I=0$ condition for the voltage lead is equivalent to $\int_\Gamma ds \, \frac{\partial \varphi}{\partial \mathbf{n}} = 0$. The solution, $\varphi(\mathbf{r})$, must satisfy this condition at the four voltage leads, with the potential at the two current leads fixed.

To satisfy all of these conditions, we take advantage of the linearity and homogeneity of Laplace's equation. If $\varphi_i (\mathbf{r})$ is a (dimensionless) solution for which the potential at contact $i$ (using the contact labeling scheme in Fig.\ \ref{figs1}(a)) is $V_i = $ 1 and $V_j = $ 0 for $j \neq i$, then for any configuration of voltages at the contacts, the full solution for the Hall bar is $\varphi (\mathbf{r}) = \sum_i V_i \varphi_i (\mathbf{r})$. It follows from this and \eqref{eq:current} that if we define a matrix $G$ by $G_{ij} = \sigma \int_{\Gamma_i} ds \, \frac{\partial \varphi_j}{\partial \mathbf{n}}$, where $\Gamma_i$ is the interface with contact $i$, then the current flowing from contact $i$ into the device is $I_i = \sum_j G_{ij} V_j$.

This equation must be inverted to find voltages that satisfy the zero current condition for the voltage leads. However, the matrix $G$ is singular due to the freedom to offset all the voltages by a fixed constant without changing the currents. We can set the potential to zero at a particular contact $d$, which we take to be the drain terminal. Then the submatrix $\widetilde{G}$ of $G$ with the $d$-th row and column removed is invertible, and $V_i = \sum_{j\neq d} \widetilde{G}^{-1}_{ij} I_j$ for $i\neq d$. Letting $I_i = 0$ for voltage leads $i$, and fixing the current $I_s$ at the source terminal $s$, the voltages (and correspondingly the full solution $\varphi$) are determined by $V_i = \widetilde{G}^{-1}_{is} I_s$. The number of squares $N_\square$ between longitudinal voltage leads $i_1$ and $i_2$, in this configuration, is then given by
\begin{equation*}
N_\square = \sigma R_{xx} = \sigma \frac{V_{i_1}-V_{i_2}}{I_s} = \sigma \left(\widetilde{G}^{-1}_{i_1 s} - \widetilde{G}^{-1}_{i_2 s} \right).
\end{equation*}

To apply this calculation to our device, we construct a polygon model of the Hall bar using an optical micrograph. We carry out the calculations in MATLAB, and numerically solve for the functions $\phi_i(\mathbf{r})$ on a triangular mesh using the built-in Partial Differential Equation Toolbox. We compute $N_\square = $ 2.77, with the full solution $\varphi(\mathbf{r})$ shown in Fig.\ \ref{figs3}. $R_{xx}$ is measured between the top two voltage leads as shown in the figure. We also find 0.026 squares between the top-right and bottom-right voltage leads used to measure $\rho_{yx}$, suggesting roughly 2.6\% mixing of $\rho_{xx}$ into the measured value of $\rho_{yx}$, in reasonable agreement with electrical measurements (see next section).\\\\

\begin{figure}
\includegraphics[width=\columnwidth]{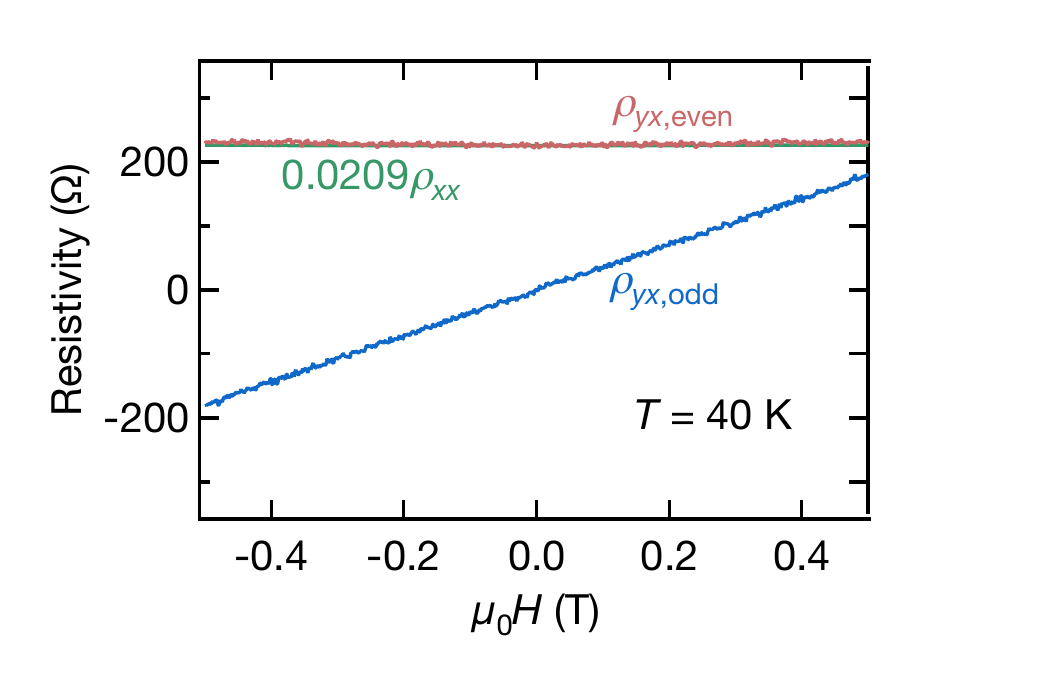}
\caption{\label{figs4} Extraction of geometric correction factor. Odd and even components of $\rho_{yx}$ measured at 40 K, above the Curie temperature. The odd components is assumed to be the true value of $\rho_{yx}$, while the even component is a linear pickup of $\rho_{xx}$, with coefficient shown here to be 2.09\%.}
\end{figure}

\begin{figure*}
\includegraphics[width=6.5in]{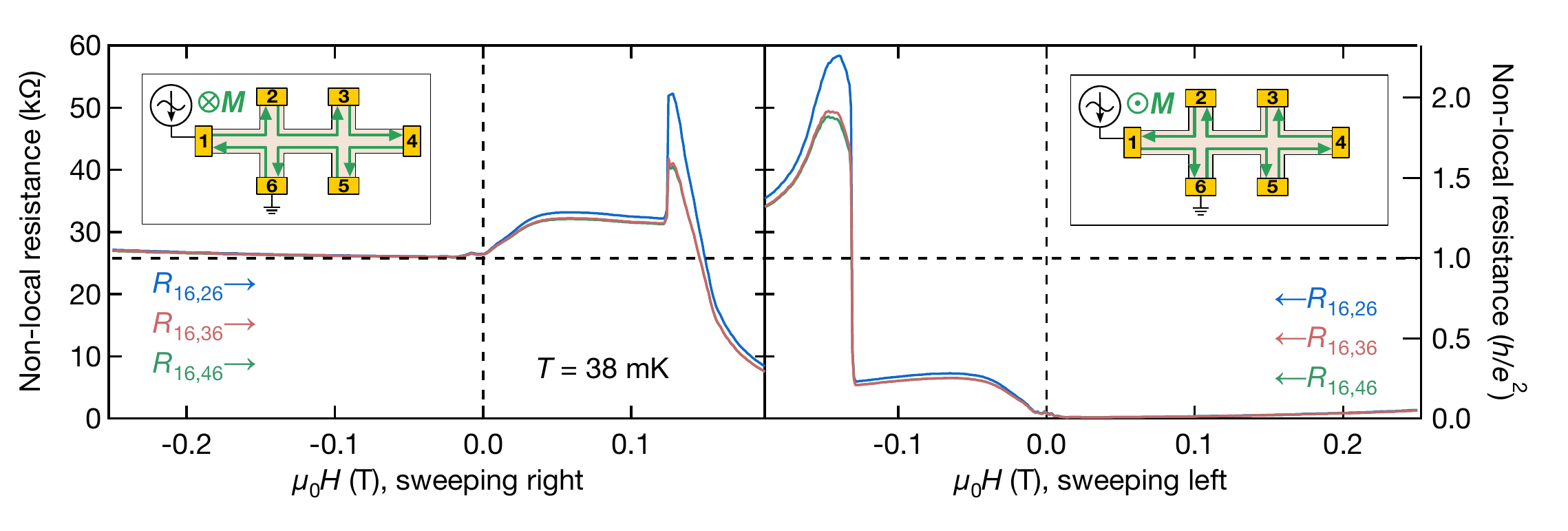}
\caption{\label{figs5} Non-local resistance measurements. Three-terminal non-local measurements, similar to Fig. 1c, but using different pairs of voltage leads. The insets show the chiralities and terminal naming scheme. The overall behavior ($R\approx h/e^2$ at $M_z=-$1, $R\approx$ 0 at $M_z=+$ 1) does not depend on choice of leads.}
\end{figure*}

\begin{figure*}
\includegraphics[width=6.5in]{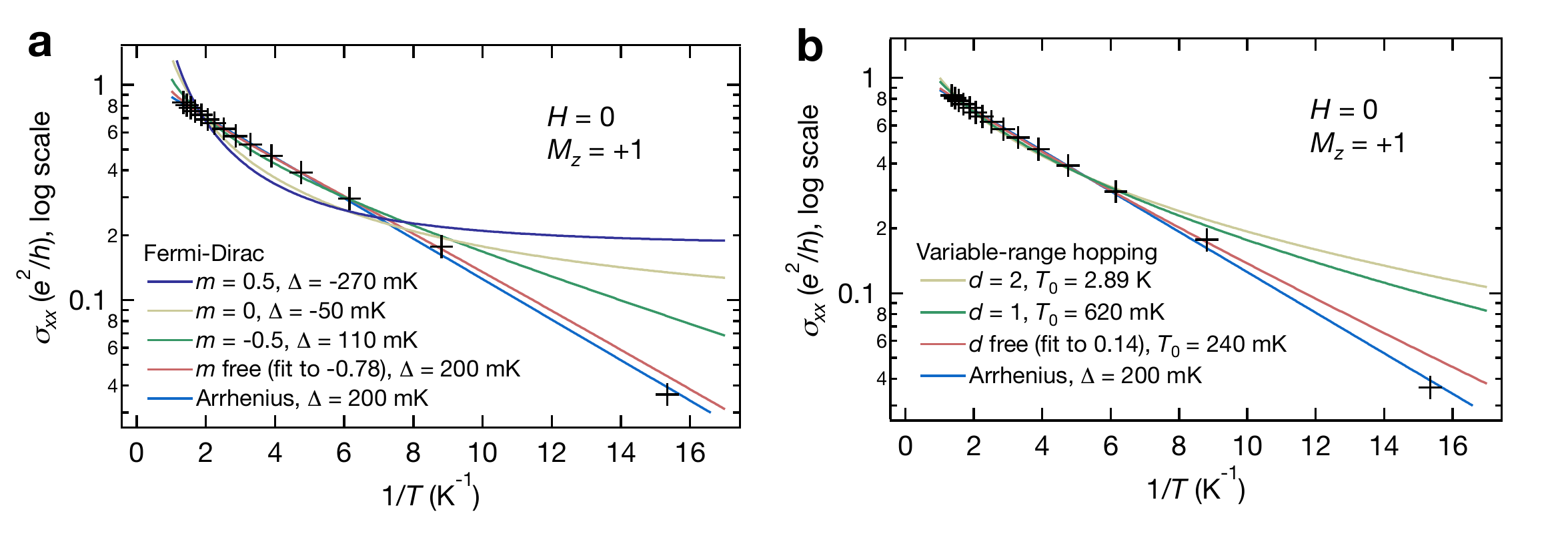}
\caption{\label{figs6} Fits of temperature dependence. (a), (b), $\sigma_{xx}$ as a function of inverse temperature, with various fits of (a) the Fermi-Dirac functional form in equation \eqref{eq:fermidirac} and (b) variable-range hopping in the form of equation \eqref{eq:vrh}. For comparison, the Arrhenius fit from Fig. 3b is included in each.}
\end{figure*}

\subsection{Correction for imperfect geometry}

A common practice for eliminating mixing between longitudinal and transverse resistances in magnetic field sweeps is to symmetrize even signals ($\rho_{xx}$) and antisymmetrize odd signals ($\rho_{yx}$) about zero field. In a system with ferromagnetic hysteresis one must perform the transformations by combining data from opposite sweep directions. However, for this device all magneto transport is history-dependent. For instance, a small difference in the sweep rate or starting value of the electron temperature can change the trajectories of otherwise identical measurements. Hence, rather than antisymmetrize (and risk combining data taken at potentially different temperatures), to remove the even component of $\rho_{yx}$ due to uneven lead spacing we subtract from it a small multiple of $\rho_{xx}$.

The value of this multiple was obtained by warming the sample to 40 K, well above the Curie temperature of its ferromagnetism, and measuring the device in a regime where symmetrizing and antisymmetrizing make logical sense. We decompose a $\rho_{yx}$ field sweep into its odd and even components (Fig.\ \ref{figs4}). At zero field, the ratio of $\rho_{yx,\text{even}}$ to $\rho_{xx}$ is 0.0209. This coefficient  works reasonably well over the full field range (Fig.\ \ref{figs4}), is in good agreement with the above Poisson calculations, and nearly eliminates asymmetries between the two magnetizations at base temperature (Figs.\ \ref{fig2}(b)-(c)). For all  data other than in Figs.\ \ref{fig1}(a), \ref{fig1}(b), and \ref{fig2}(b), we report $\rho_{yx} \equiv $$\rho_{yx\text{,measured}}-\text{0.0209}\rho_{xx}$.\

Note that in this regime, $\rho_{yx,\text{even}}\sim\rho_{yx,\text{odd}}$ even though the geometric imperfection is small. This is due to low mobility and low magnetic field. Of course at quantization $\rho_{yx}$ is dominated by the Hall effect, but the mixing correction remains important at the level of precision we seek.

\subsection{Non-local measurements}

Additional non-local measurements are shown in Fig.\ \ref{figs5}. Sourcing current from terminal 1 and draining at 6, we measure three-terminal resistances between the drain and contacts 2, 3, and 4 as the applied field is swept (Fig.\ \ref{figs5}(a)). All three measurements show similar behavior to Fig.\ \ref{fig1}(c): when the magnetization is negative, the voltages equilibrate in a clockwise manner, so $R_{16,26} \approx R_{16,36} \approx R_{16,46} \approx h/e^2$; with positive magnetization, the equilibration is reversed, and $R_{16,26} \approx R_{16,36} \approx R_{16,46} \approx 0$. Away from $|H|\approx$ 0, the same activated conduction discussed in the main text causes deviations from these values. The small differences between the three measurements can be attributed to the contact resistance to the drain and possibly, near quantization, effects of other conduction paths including residual regions of surface conduction and non-chiral edge conduction. These conduction channels are best characterized with four-terminal non-local measurements between adjacent pairs of contacts, as performed previously~\citeS{SUCLA}.

\subsection{Fits of temperature dependence}

It is somewhat surprising that $\sigma_{xx}$ as a function of inverse temperature is fit well by a simple exponential. Thermal activation over a gap into a single band typically has the form
\begin{equation*}
\sigma_{xx} \propto \int_{E_{F}}^{\infty}\mu(E)D(E)\frac{1}{1+e^{(E-E_{F})/k_BT}}dE
\end{equation*}
where $E_F$ is the Fermi energy, $\mu$ is the carrier mobility as a function of energy, $D$ is the density of states (zero in the band gap and with some dependence on $E$ in the conduction band), and the fraction is the Fermi-Dirac function. If we assume that the Fermi level is energy $\Delta$ below the bottom of the conduction band and the product $\mu(E)D(E)$ has a simple power law form $aE^m$ (where we set $E=0$ at the band bottom), then this integral becomes
\begin{equation}\label{eq:fermidirac}
\sigma_{xx} \propto \int_{\Delta}^{\infty}\frac{aE^m}{1+e^{E/k_BT}}dE.
\end{equation}
Leaving $a$ and $\Delta$ as free parameters and using generic values of $m$ (namely, 0 and $\pm$1/2, resulting from constant $\mu$ and either linear or quadratic dispersions in 1D or 2D), fits of this integral to the data do not match (Fig.\ \ref{figs6}(a)). If we let $m$ also be a free parameter, then we achieve a reasonable fit over the full temperature range with $m$ = 0.78 (Fig.\ \ref{figs6}(a), red line), though without an obvious physical interpretation. Meanwhile, the simple Arrhenius fit continues to work just as well.

In quantum Hall systems, the justification for using the Arrhenius law is that $D(E)$ can be treated as a delta function at the energy of the nearest Landau level, and the Fermi-Dirac function simplifies to a Boltzmann distribution in the limit of $k_BT\ll\Delta$. The latter condition appears to be routinely relaxed in published data, where Arrhenius fits often hold up to $k_BT = 2\Delta$~\citeS{SArrhenius1, SArrhenius2}. In our case, it appears to hold even higher, up to $4\Delta$. Although we believe the exponential form is a strong indication that thermal activation is taking place, it is unclear whether this continued trend at high temperature is a coincidence resulting from the particular mobility-density of states product or the consequence of a deeper physical principle about the material or the QAHE.

In a previous QAHE study~\citeS{STsinghua}, the relationship between $\sigma_{xx}$ and $\sigma_{xy}$ suggested that bulk conduction arose from variable-range hopping. In that were the case in our sample, the temperature dependence would have the form
\begin{equation}\label{eq:vrh}
\sigma_{xx} \propto e^{-(T_0/T)^{1/(d+1)}}
\end{equation}
where $d$ is the number of spatial dimensions of the system and $T_0$ is a temperature scale. We attempted to fit this function to the data for $d=$ 2, $d=$ 1, and letting $d$ be a free parameter (Fig.\ \ref{figs6}(b)). The first two cases fit poorly while the third fits $d$ to 0.14 (i.e., close to reducing to the Arrhenius function). We therefore believe that variable-range hopping is not an accurate description of the temperature dependence of this device.

\subsection{Heating events in field sweeps}

Throughout this paper we argue that -- below the coercive field -- the variations in transport with applied field in our magnetic TI device can largely be explained by temperature variations due to thermal contact with some (de)magnetizing system. (For further analysis about the nature of that system, see section below). In addition to this overall pattern, we observe other features in $\sigma_{xx}$ that presumably arise from heating. Namely, while sweeping the applied field toward zero, $\sigma_{xx}$ (and therefore the inferred temperature) briefly increases around $|\mu_0H|=$ 20 mT and also more dramatically right before reaching zero (see Fig.\ \ref{fig2}(a)). The latter was observed by some of us previously~\citeS{SUCLA}. The magnitude of these effects depends strongly on the magnet ramp rate: in slower sweeps they are far less prominent, and they tend to dissipate quickly when holding field constant (see the quick drop in $\rho_{xx}$ between the third and fourth panels of Fig.\ \ref{fig4}(c)). We do not know the specific mechanism for either feature, but these shorter timescales suggest that the heating occurs closer to the dilution refrigerator's mixing chamber.

\subsection{Analysis of demagnetization}

\begin{figure*}
\includegraphics[width=6.5in]{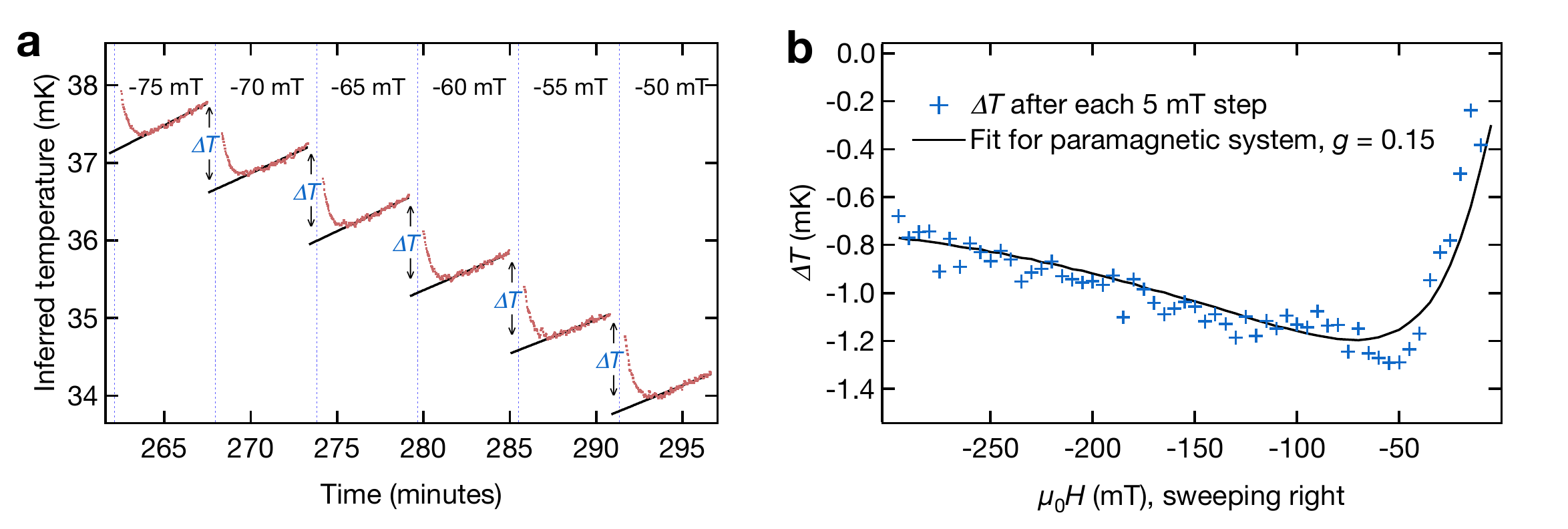}
\caption{\label{figs7} Fit of magnetocaloric effect. (a), Extraction of the drop in inferred temperature for each 5 mT field step, removing the effect of thermal equilibration with the cryostat. Note that data were only taken during the five-minute waits between 50-second field steps. (b), Inferred temperature drop at each step, with a fit using \eqref{eq:entropy}.}
\end{figure*}

In this section we discuss the possible sources of the observed magnetocaloric effect. A first candidate is the sample's magnetism itself. The sharpness of the anomalous Hall hysteresis loop indicates that, overall, it is a hard ferromagnet and therefore not experiencing major changes in entropy except at transitions. However, if the magnetism is spatially inhomogeneous, there may be isolated weak patches with softer ferromagnetism and a magnetocaloric effect without contributing to transport measurements. Another possibility is that clusters of chromium atoms may exhibit paramagnetism or weak ferromagnetism~\citeS{Sclusters}. In both cases, we would expect the sign of the effect to change as the total magnetic field, $B$ (which includes the sample magnetization), crosses zero. Instead, it responds to $H$, suggesting that the demagnetizing system is a least somewhat spatially separated from the electron system being measured. In the immediate vicinity of the TI film are the semi-insulating GaAs substrate and the indium contacts. To pin the Fermi level in its band gap, the semi-insulating GaAs wafer is doped with carbon (rather than chromium, as is sometimes used), but perhaps some remanent magnetic impurities are nonetheless present, and even carbon doping has been associated with paramagnetism in GaAs~\citeS{SGaAspara}. The indium metal could also contain magnetic impurities.

For now, we assume that one of the scenarios above holds -- the magnetic system responsible for the effect is located near the device -- and attempt to extract some information about that magnetic system based on our measurements. (We later consider the possibility that this magnetic system may be located elsewhere in the cryostat, and explain why we consider it unlikely.) We make several other simplifying assumptions: 1) that the relevant field scale $B$ for the magnetocaloric effect is equal to the applied field $\mu_0 H$ (an underestimate if the film's magnetization contributes to $B$); 2) that the magnetocaloric system is in good thermal contact with a constant-in-temperature heat capacity $C_V$ that includes the TI film; 3) that we can, under the right circumstances, model the demagnetization as adiabatic; 4) that we can treat the magnetocaloric system as a system of $N$ non-interacting spins, each of which has energy $\pm g\mu_{0}B$ (where $g$ is the Land\'e g-factor and $\mu_{0}$ is the Bohr magneton). The latter assumption leads to canonical partition function
\begin{equation*}
Z = \left(2 \cosh{\frac{g\mu_{0}B}{k_{B}T}}\right)^N
\end{equation*}
where $k_{B}$ is the Boltmann constant and $T$ is the temperature. Then the entropy is
\begin{widetext}
\begin{equation*}
S = \frac{\partial}{\partial T}\left(k_{B}T\ln{Z}\right) = N k_{B} \left[\ln\left({2\cosh{\frac{g\mu_{0}B}{k_{B}T}}}\right) - \frac{g\mu_{0}B}{k_{B}T}\tanh{\frac{g\mu_{0}B}{k_{B}T}}\right]
\end{equation*}
To put this system in good thermal contact with the electronic system, we add $C_{V}\ln{T/T_0}$ to the entropy, where $T_0$ is an arbitrarily low temperature. In any adiabatic process the total entropy does not change, and so the  quantity
\begin{equation}\label{eq:entropy}
k_{B} \frac{N}{C_{V}}\left[\ln\left({2\cosh{\frac{g\mu_{0}B}{k_{B}T}}}\right) - \frac{g\mu_{0}B}{k_{B}T}\tanh{\frac{g\mu_{0}B}{k_{B}T}}\right] + \ln{T/T_0}
\end{equation}
\end{widetext}
is held constant. If we change $B$ adiabatically, then $T$ will adjust accordingly. In a fit a measured trajectory $T(B)$, the only two free parameters are $g$ and $N/C_{V}$ ($T_0$ only adds a constant offset), where, roughly speaking, $N/C_{V}$ determines the size of the effect and $g$ determines the field scale at which it takes place.

In practice, adiabaticity is a poor assumption because the sample exchanges heat with the cryostat. However, the data in Figure \ref{fig4}(a), when converted to temperature, can provide the rate of this heat exchange, allowing us to approximately isolate the effect of demagnetization. We do this by performing a linear fit to the temperature drift during the waiting periods and extrapolating it back to the start of the field step (Fig.\ \ref{figs7}(a)). The result provides what the approximate change in temperature for each 5 mT field step would be if it were adiabatic, which we can fit with the simple model above (Fig.\ \ref{figs7}(b)). We find that the data best match the model with a g-factor of $\sim$0.15 and $N/C_V=\text{1.7}\times\text{10}^{\text{24}}\text{ K}/\text{J}=\text{23}/k_B$. However, a more realistic choice of heat capacity would likely decrease at lower temperatures. The result would be a larger drop in temperature at the rightmost region of Fig.\ \ref{figs7}(b), which the fit would correct for by reducing the value of the g-factor. (On the other hand, if the magnetocaloric system is located outside the magnet bore and hence experiences a field smaller than $\mu_0H$, the estimate for $g$ would need to increase.)

For an isolated semi-insulating GaAs sample, a relatively small concentration of magnetic moments could lead to a magnetocaloric effect on the order of what we observe. Using a simple Debye model calculation (and the known GaAs Debye temperature $T_D = $345 K in the low temperature limit~\citeS{Sdebye}), the heat capacity per atom is
\begin{equation*}
\frac{C_V}{N_\text{atoms}} = \frac{\text{12}\pi^4}{\text{5}}k_B\left(\frac{T}{T_D}\right)^3\approx(\text{4}\times\text{10}^{-\text{10}})k_B\text{ at 40 mK}
\end{equation*}
In our fit to the magnetocaloric effect, we find $N/C_V=\text{23}/k_B$, which implies a concentration $N/N_\text{atoms}\approx\text{10}^{-\text{8}}$, placing $N$ in the low 10$^\text{14}\text{ cm}^{-\text{3}}$ range. This number, low enough to be a reasonable impurity concentration, is a consequence of the small heat capacity of the insulating crystal. In reality, our system is also comprised of the TI film, indium contacts, wires, and eventually the rest of the cryostat. But if the film has an insulating surface and bulk, the interior of the device could become a poor thermal conductor and isolate the rest of those elements from much of the cold substrate.

The timescales of the magnetocaloric effect support this picture. As Fig.\ \ref{fig2}(d) demonstrates, $\rho_{xx}$ (and therefore, we believe, the temperature of the film) can stay well below the equilibrium value for hours. After a field sweep down, the cooling also continues for tens of seconds (see Fig.\ \ref{figs7}(a)), a timescale that lengthens to minutes at the very lowest temperatures. Both of these details make sense if the demagnetizing process takes place at or next to the surface of the film, which itself becomes increasingly thermally resistive as it cools. The latter detail is even better explained if the cooling is spatially inhomogeneous, so that cooler and warmer parts of the film take some time to equilibrate.

\begin{figure*}
\includegraphics[width=6.5in]{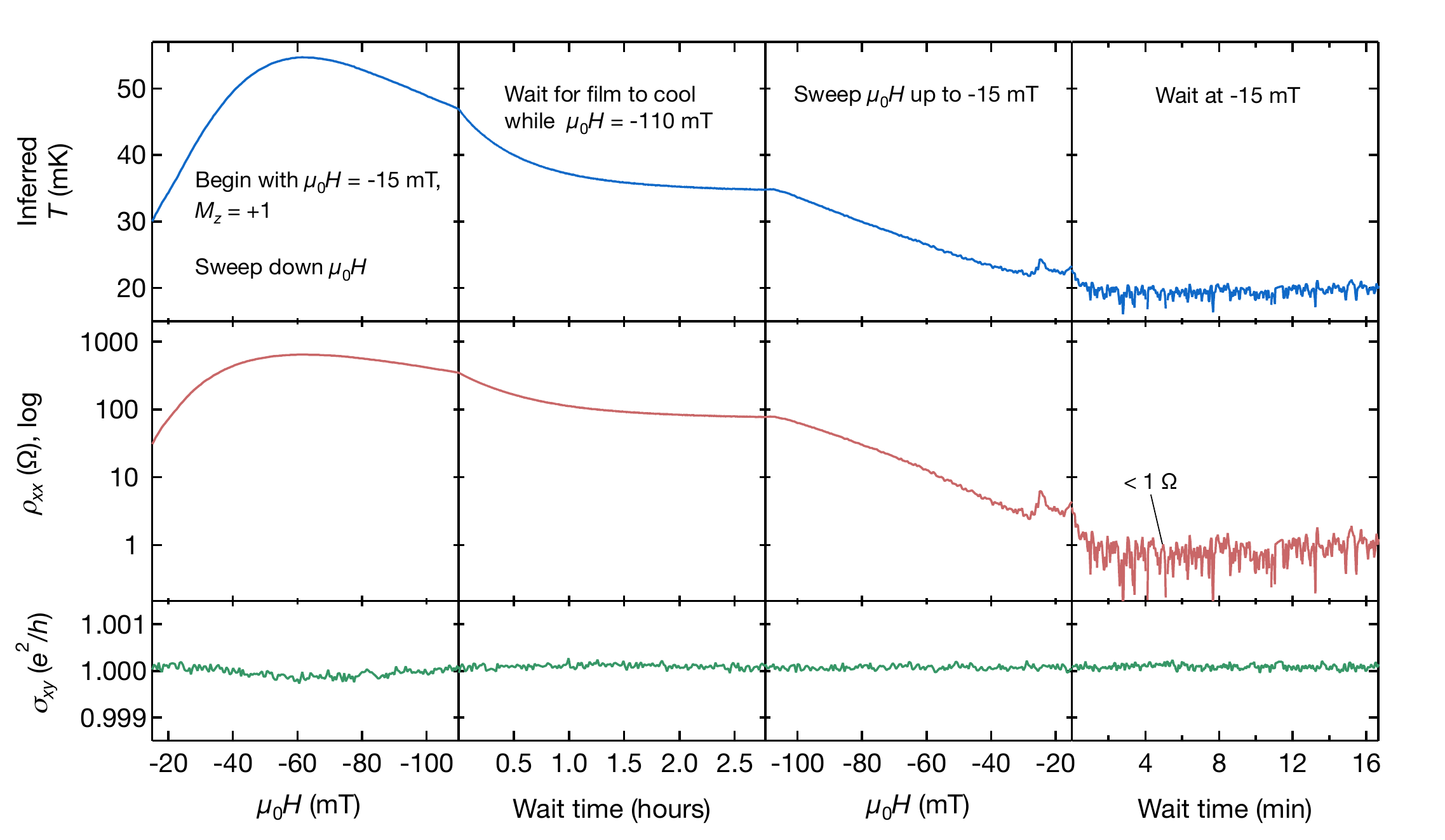}
\caption{\label{figs8} Adiabatic demagnetization cooling cycle. $\rho_{xx}$, $\sigma_{xy}$, and the temperature of the film, as inferred from the Arrhenius fit, are shown for an adiabatic demagnetization cycle. Beginning at $H=-15$ mT, the applied field is first swept slowly to -110 mT, after which the film is given time to cool. The field is then swept quickly back to -15 mT, causing demagnetization cooling that brings the apparent temperature of the film near 20 mK. The film cools further upon waiting several minutes, with $\rho_{xx}$ dropping below 1 $\Omega$.}
\end{figure*}

\begin{figure*}
\includegraphics[width=6.5in]{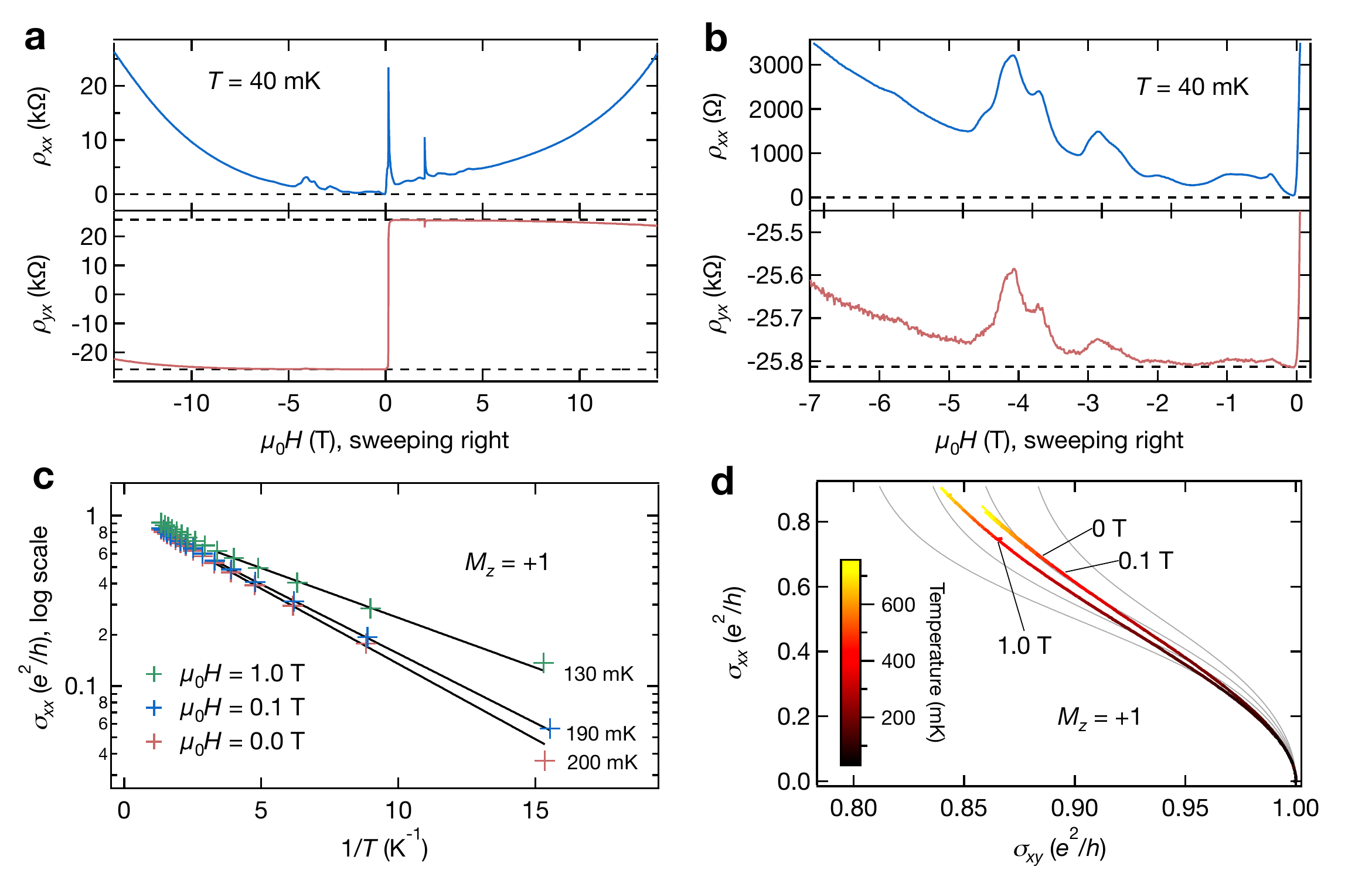}
\caption{\label{figs9} High field measurements. (a), Longitudinal and Hall resistivities during field sweep over magnet's full range, -14 T to 14 T. (b), Smaller range of same data, showing resistivities as applied field approaches and crosses zero. (c), Arrhenius plots of $\sigma_{xx}$ versus inverse temperature at three applied fields. The lines are exponential fits with characteristic temperatures as labelled. (d), Parametric plot of $\sigma_{xx}$ versus $\sigma_{xy}$ as the sample is warmed at three different applied field. The gray lines are calculated renormalization group flow lines. The plots for 0 T and 0.1 T applied fields are indistinguishable.}
\end{figure*}

In light of this information, let us summarize the various possibilities previously identified. The indium contacts are probably too easily thermalized to be a candidate, and the low extracted g-factor is probably inconsistent with the Cr cluster picture. The above analysis does support the possibility of a small amount of magnetism in the GaAs substrate, which requires a device with low heat capacity and thermal conductivity to be relevant. The scenario with patches of weak magnetism is also consistent with experiment details, but probably not well described by the simple paramagnet model above.

The other main class of scenarios involve having a paramagnetic system located elsewhere in the cryostat. We have noticed a magnetocaloric effect in measurements of this type of magnetically-doped film at both Stanford and (in retrospect) UCLA~\citeS{SUCLA}. It is conceivable that both cryostats contain paramagnetic materials that give rise to this effect. During two years of measurements in our cryostat at Stanford, magnetically-doped TI films grown at UCLA are the only samples found to show magnetocaloric effect, but many other samples do not have strongly temperature-sensitive transport below 100 mK, so the effect might be missed elsewhere.

Some experimental details put constraints on where such a magnetic system may be located. The magnet is compensated so that its field at the mixing chamber is a factor of 500 lower than at the center of the bore. If the magnetic system were in such a low field region, its g-factor would have to be extremely large ($\sim$100) to match the measured field response. We therefore believe any candidate must be in the bottom 10 cm or so of the probe's cold finger. Its thermal conduction to the sample must be much better than to the probe's mixing chamber stage, where thermometry measurements do not show any response to demagnetizing. However, the cold finger, made of oxygen-free high conductivity copper, is designed to fully thermalize with the mixing chamber in order to maximally cool the sample at its bottom. If such a location existed, with poor conduction to the cryostat but not sample, it would be likely to act as a weak link, impairing cooling power to the sample and raising its base temperature.

\subsection{Alternative demagnetization cycle}

To achieve the lowest possible $\rho_{xx}$, a sequence similar to an adiabatic demagnetization cycle can be used. We can further optimize from the cycle used in the main text by ending at a small but nonzero applied field. This allows us to avoid the heating event near zero field, and also to stop before the cooling power from adiabatic demagnetization tends toward zero. This can be seen in Fig.\ \ref{figs7}(b). In the simple model developed above, the quantity \eqref{eq:entropy} remains constant during an adiabatic process. Differentiating with respect to $B$, it follows that
\begin{widetext}
\begin{equation}\label{eq:dT/dB}
\frac{dT}{dB} = \frac{\frac{g \mu_0 N}{C_V} \left( \frac{g \mu_0 B}{k_B T}\right) \sech^2\left( \frac{g \mu_0 B}{k_B T}\right)}{\frac{k_B N}{C_V} \left( \frac{g \mu_0 B}{k_B T}\right)^2 \sech^2\left( \frac{g \mu_0 B}{k_B T}\right) + 1}
= \frac{g \mu_0 N}{C_V} \left(\frac{g \mu_0 B}{k_B T}\right) + \mathcal{O}\left(\left(\frac{g \mu_0 B}{k_B T}\right)^2\right).
\end{equation}
\end{widetext}
Since $T$ remains nonzero, $dT/dB \to$ 0 as $B$ approaches zero, so it is advantageous to stop the adiabatic demagnetization at a nonzero field before the heat load overwhelms the cooling power.

We performed a demagnetization cycle beginning and ending at $H = -$15 mT and report  $\rho_{xx}$, $\sigma_{xy}$, and the inferred temperature from the Arrhenius fit (Fig.\ \ref{figs8}). Following a previous demagnetization, the cycle starts with the film at an inferred temperature of 30 mK, still below the equilibrium temperature of 35 mK. In a rough approximation of isothermal magnetization, the magnitude of the applied field is then slowly increased until $H = -$110 mT, where the film is allowed to cool after heating from the magnetization. Here, the field is changed at an average rate of about 0.011 mT/s, roughly 20\% slower than the corresponding step in the cycle described in the main text in order to decrease the amount of heating. The adiabatic demagnetization step is then performed by sweeping $H$ from -110 mT to -15 mT at an average rate of 0.042 mT/s. By comparison, the corresponding step in cycle from the main text was performed at approximately the same rate from -110 mT to -30 mT, and then at 0.013 mT/s from -30 mT to 0. After reaching -15 mT, $\rho_{xx}$ is around 4 $\Omega$. However, upon waiting a short time, the film cools further and $\rho_{xx}$ drops below 1 $\Omega$.

\begin{figure*}
\includegraphics[width=6.5in]{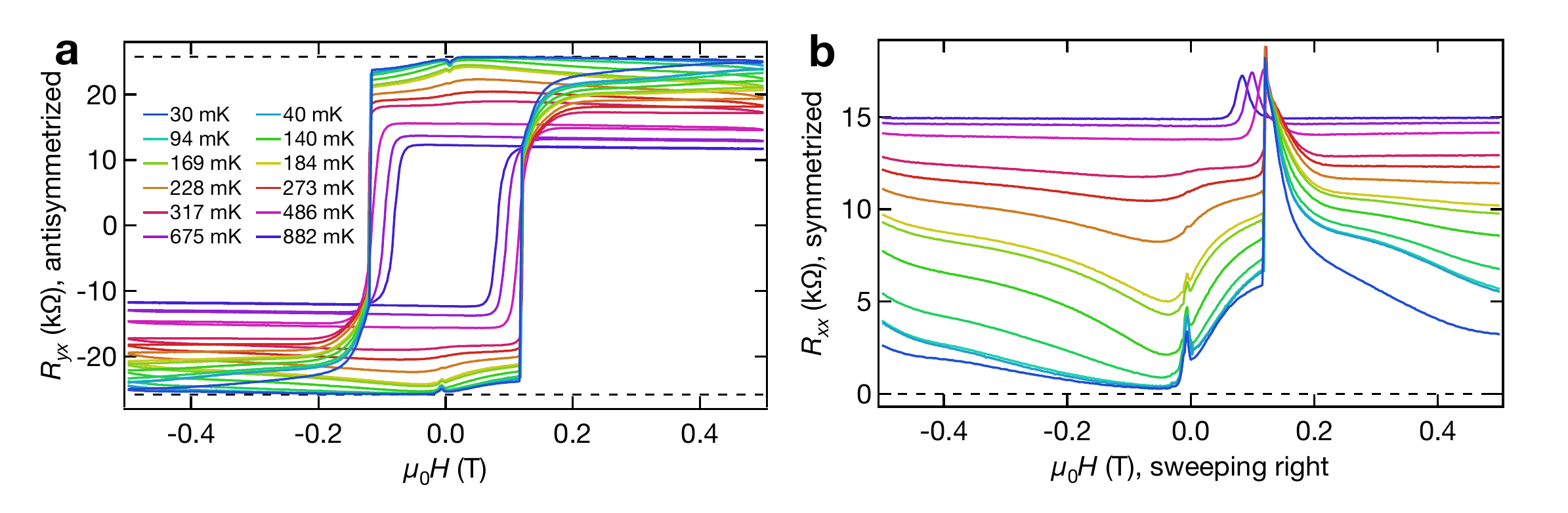}
\caption{\label{figs10} Hysteresis loops at higher temperatures. (a), Hall resistance of a different sample made from the same material, in ferromagnetic hysteresis loops, at a range of temperatures. The calibration in these data is less precise and the magnet sweep rate is much faster (1 mT/second, except 30 mK sweeps, which are 0.4 mT/second) than for all other 
figures. (b), Longitudinal resistance during same measurements. For clarity, only half of the loop is shown.}
\end{figure*}

\subsection{High-field measurements}

Although the main emphasis of this paper is the QAHE at low applied fields, the system has surprising behavior at higher fields. A sweep over the magnet's full field range, from -14 T to 14 T (Fig.\ \ref{figs9}(a)), demonstrates that $\rho_{xx}$ becomes sizable and $\rho_{yx}$ departs substantially from quantization above $\pm$3 T. This was not the case in previous measurements performed by some of us~\citeS{SUCLA}, but we have in several cases observed that as samples age they become more resistive at high field. (This effect is even more acute in samples without good quantization.) The device measured in this paper was made from a film that had been grown six months earlier.

Some large resistivity features appear during the sweep, such as between -5 T and -2T (Fig.\ \ref{figs9}(b)). Under the main interpretation proposed in this paper, these would be seen as heating events, but they persist over many hours of sweeping time. The feature near +2 T, on the other hand, coincides with a temporary spike in the temperature of the mixing chamber plate thermometer.

When repeating the high-temperature measurements at nonzero applied field, the energy scale of the thermal activation decreases (Fig.\ \ref{figs9}(c)), suggesting a closing exchange gap. The system's trajectory in conductivity space shifts as well between 0 and 1 T (Fig.\ \ref{figs9}(d)), though the 0.1 T trajectory is indistinguishable from that at zero field.

\subsection{Hysteresis loops at higher temperatures}

In a previous device made from the same material (also different from that in Kou \emph{et al.}~\citeS{SUCLA}), hysteresis loops were taken at a variety of temperatures (Fig.\ \ref{figs10}(a)-(b)). These measurements were less careful: the calibration was performed to a known resistor rather than a more precise quantum Hall plateau, the lock-in amplifier's excitation frequency was higher (leading to substantial phase offsets between excitation and measurement), and the magnetic field sweep rate was much faster (leading to some heating in cryostat elements due to eddy currents). Nonetheless, they add useful information to our understanding of the ferromagnetism in the film.

As is the case elsewhere, raising the temperature to 100 mK quickly increases the value of $R_{xx}$ and degrades quantization in $R_{yx}$. Between 317 mK and 486 mK we observe several changes: the hysteresis loop becomes much less sharp, the coercive field starts to decrease, and the demagnetization effect is no longer visible in $R_{xx}$. Further study is warranted in this temperature range to understand if these are related effects.

\bibliographystyleS{apsrev4-1}
\bibliographyS{QAHE}

\end{document}